\documentclass[a4paper,11pt]{article}
\pdfoutput=1 
\usepackage{jheppub} 
\usepackage{bm}
\usepackage[T1]{fontenc} 

\newcommand{\ri}{{\rm i}}

\preprint{EPHOU-23-014}

\title{\boldmath Moduli trapping mechanism in modular flavor symmetric models}

\author[a]{Shota Kikuchi,}
\author[a]{Tatsuo Kobayashi,}
\author[a]{Kaito Nasu,}
\author[b]{Yusuke Yamada}

\affiliation[a]{Department of Physics, Hokkaido University, Sapporo 060-0810, Japan}
\affiliation[b]{Waseda Institute for Advanced Study, Waseda University, 1-21-1 Nishi Waseda, Shinjuku, Tokyo 169-0051, Japan}

\abstract{We discuss how the moduli in modular flavor symmetric models dynamically select enhanced symmetry points at which the residual modular symmetry renders extra matter fields massless. The moduli dynamics non-perturbatively produces the extra matter particles, which gives (time-dependent) effective potential that traps the moduli to enhanced symmetry points. We show analytic estimates of particle production rate consistent with numerical results, and the dynamics of moduli based on the analytic estimates.}

\begin{document} 
\maketitle
\flushbottom
\section{Introduction}\label{Intro}
In effective theory of superstring, moduli fields, light fields associated e.g. with higher-dimensional gravitational degrees of freedom, play crucial roles in constructing realistic models of particle physics as well as cosmology. 

Modular symmetry is the geometrical symmetry of compact space, where moduli transform 
non-trivially.
Four-dimensional effective field theory derived from superstring theory also has 
the modular symmetry \cite{Ferrara:1989bc,Ferrara:1989qb,Lerche:1989cs,Lauer:1989ax,Lauer:1990tm,Kobayashi:2018rad,Kobayashi:2018bff,Ohki:2020bpo,Kikuchi:2020frp,Kikuchi:2020nxn,Kikuchi:2021ogn,Almumin:2021fbk,Baur:2019kwi,Baur:2019iai,Nilles:2020kgo,Nilles:2020gvu}.
In addition, the modular symmetry includes finite groups such as 
$S_3$, $A_4$, $S_4$, and $A_5$ \cite{deAdelhartToorop:2011re}.
These discrete groups have been used in 
flavor model building in the bottom-up approach \cite{Altarelli:2010gt,Ishimori:2010au,Kobayashi:2022moq,Hernandez:2012ra,King:2013eh,King:2014nza,Tanimoto:2015nfa,King:2017guk,Petcov:2017ggy}.
Inspired by these aspects, modular flavor symmetric models 
have been receiving attention as an origin of flavor structure in the standard model of particle physics \cite{Feruglio:2017spp,Kobayashi:2018vbk,Penedo:2018nmg,Criado:2018thu,Kobayashi:2018scp,Novichkov:2018ovf,Novichkov:2018nkm,deAnda:2018ecu,Kobayashi:2018wkl,Okada:2018yrn,Novichkov:2018yse,Ding:2019xna,Nomura:2019jxj,Novichkov:2019sqv,Okada:2019uoy,deMedeirosVarzielas:2019cyj,Nomura:2019yft,Kobayashi:2019rzp,Liu:2019khw,Okada:2019xqk,Kobayashi:2019mna,Ding:2019zxk,King:2019vhv,Nomura:2019lnr,Criado:2019tzk,Kobayashi:2019xvz,Asaka:2019vev,Ding:2019gof,Zhang:2019ngf,Wang:2019ovr,Kobayashi:2019uyt,Nomura:2019xsb,Kobayashi:2019gtp,Lu:2019vgm,Wang:2019xbo,King:2020qaj,Abbas:2020qzc,Chen:2020udk,Okada:2020oxh,Okada:2020dmb,Ding:2020yen,Ding:2020msi,Okada:2020rjb,Novichkov:2020eep,Liu:2020akv,deMedeirosVarzielas:2020kji,Asaka:2020tmo,Okada:2020ukr,Wang:2020lxk,Okada:2020brs,Yao:2020qyy,Li:2021buv}\footnote{See for more references Ref.~\cite{Kobayashi:2023zzc}.}.

In the modular flavor symmetric models, 
Yukawa couplings correspond to modular forms of complex structure moduli $\tau$ , and the structure of Yukawa couplings is determined by the vacuum expectation values (VEVs) of complex structure moduli, which are complex scalar fields in 4D effective theory. Therefore, moduli stabilization plays a crucial role in such flavor symmetry models. Indeed, moduli stabilization has been studied in modular flavor symmetric models \cite{Kobayashi:2019xvz,Kobayashi:2019uyt,Abe:2020vmv,Ishiguro:2020tmo,Novichkov:2022wvg,Ishiguro:2022pde,Leedom:2022zdm,Knapp-Perez:2023nty}.

In particular, residual $Z_N$ symmetries remain at fixed points \cite{Novichkov:2018ovf}. Each modular form behaves like $(\tau - \tau_*)^m$ 
around a fixed point $\tau_*$, depending on the $Z_N$ charge $m$ of 
the modular form.
Thus, the modular forms are suppressed by $\varepsilon = \tau -\tau_*$ at nearby fixed point $\tau_*$.
This behavior is important to realize the hierarchy of quark and lepton masses and mixing angles \cite{Feruglio:2021dte,Novichkov:2021evw,Petcov:2022fjf,Kikuchi:2023cap,Abe:2023ilq,Kikuchi:2023jap,Abe:2023qmr,Petcov:2023vws,Abe:2023dvr}.

One of the ways to stabilize the moduli is the use of fluxes of $p$-form fields along compactified dimensions that yield effective potential to complex structure moduli. If somehow the chosen flux potential yields preferable VEV to the complex structure moduli, it would explain the flavor structure in the standard model by chance. Another possibility is due to non-perturbative effects. On the other hand, there might be some dynamical origin why the complex structure moduli may take some special value in the moduli space: In \cite{Kofman:2004yc}, it was pointed out that particle production takes place when moduli dynamically crosses the special point called the enhanced symmetry point (ESP) at which particles coupled to the moduli become massless. The produced particle gives effective potential to the moduli such that the moduli are attracted to the ESP and trapped around it. Such mechanism may explain why the VEV of moduli chooses some particular value from the landscape of moduli space.

In this work, we consider the dynamical moduli trapping mechanism in modular flavor symmetric models. First, with simple toy models we briefly review and develop a numerical approach to discuss the moduli dynamics back-reacted by the production of spectator fields. Despite its straightforwardness, we find a common difficulty due to the computational costs in the simulations. Furthermore, the modular flavor symmetric models contain extra complications due to the structure of Yukawa couplings as well as moduli being coupled to matters via Planck suppressed operators. Therefore, we will develop a semi-analytic approach where only the first particle production event is taken into account, and contributions from all the momentum modes are included. Using the analytic result, we show that the moduli trapping mechanism works even for (complicated) modular flavor symmetric models.

This paper is organized as follows. In Sec.~\ref{toymodel}, we review and discuss a fully numerical approach, which is in principle applicable to any dynamical systems. In Sec.~\ref{Modularflavor}, we review the modular flavor symmetry , and then specify the system we consider in this work and derive the equations of motion that we need to solve. We then analytically evaluate the amount of particle production when moduli crosses the ESP in Sec.~\ref{analyticPP}. We also numerically examine the moduli dynamics using the analytic expressions we derived. Finally, we summarize our results and discuss implications to more realistic models in Sec.~\ref{summary}.
In Appendix \ref{PPreview}, we give a brief review on particle productions 
in time-dependent background.
Modular forms, which we use, are shown in Appendix \ref{modular-form}.
In Appendix \ref{freedynamics}, we review field dynamics in expanding Universe.

Throughout this paper, we take a natural unit convention $\hbar=1$, $c=1$ but we write the Planck scale $M_{\rm pl}\sim 2.4\times 10^{18}$GeV explicitly to clarify the hierarchy between various scales.

\section{A toy model}\label{toymodel}
We briefly review some ingredients that we will use in modular flavor symmetric models. For simplicity, we discuss the following system consists of two massless real scalar fields,
\begin{align}
    S=-\frac12 \int d^4x\sqrt{-g}\left[(\partial\phi)^2+(\partial\chi)^2+\lambda\phi^2\chi^2\right].
\end{align}
We assume that $\phi$ behaves as a classical homogeneous field $\phi=\phi(t)$, the background to be the Friedman-Robertson-Walker (FRW) spacetime $ds^2=-dt^2+a^2(t)d{\bm x}^2$ and $\chi$ to be a quantum field, and then the coupling term $\lambda\phi(t)^2\chi^2$ behaves as an effective time-dependent mass term of $\chi$. We quantize $\chi$ as follows: First, the action of $\chi$ is
\begin{equation}
    S=\frac12 \int dt d^3xa^3\left[\dot{\chi}^2-a^{-2}(\partial_i\chi)^2-\lambda\phi^2\chi^2\right],
\end{equation}
which, by introducing $\tilde{\chi}=a^{\frac32}\chi$, can be rewritten as
\begin{align}
    S=&\frac12 \int dt d^3x\left[\left(\dot{\tilde{\chi}}-\frac32H\tilde{\chi}\right)^2-a^{-2}(\partial_i\tilde{\chi})^2-\lambda\phi^2\tilde{\chi}^2\right]\nonumber\\
    =&\frac12 \int dt d^3x\left[\dot{\tilde{\chi}}^2-a^{-2}(\partial_i\tilde{\chi})^2-\left(\lambda\phi^2-\frac{3}{2}\dot{H}-\frac{9}{4}H^2\right)\tilde{\chi}^2\right].
\end{align}
Then, the quantum field $\hat{\chi}$ is written by 
\begin{equation}
    \hat{\chi}(t,{\bm x})=\frac{1}{a^{\frac32}(t)}\int \frac{d^3k}{(2\pi)^{\frac32}}\left[\hat{a}_{\bm k}e^{\ri {\bm k}\cdot{\bm x}}f_k(t)+\hat{a}^{\dagger}_{\bm k}e^{-\ri {\bm k}\cdot{\bm x}}f^*_k(t)\right],
\end{equation}
where the mode function $f_k(t)$ satisfies
\begin{equation}
    \ddot{f}_k(t)+\omega_k^2(t)f_k(t)=0,\label{modeeq}
\end{equation}
and
\begin{equation}
    \omega_k^2(t)=\frac{k^2}{a^2(t)}+\lambda\phi^2(t)-\frac32 \dot{H}(t)-\frac94 H^2(t).
\end{equation}
The mode function $f_k(t)$ satisfies the normalization condition
\begin{equation}
    f_k\dot{f}_k^*(t)-f^*_k(t)\dot{f}_k(t)=\ri,\label{KKnorm}
\end{equation}
which implies the canonical commutation relation
\begin{equation}
    [\hat{a}_{\bm k},\hat{a}^{\dagger}_{{\bm k}'}]=\delta^3({\bm k}-{\bm k}') \ \text{or equivalently} \ [\hat{\chi}(t,{\bm x}),\dot{\hat{\chi}}(t,{\bm y})]=\ri \delta^3({\bm x}-{\bm y}),\label{CCR}
\end{equation}
for all $t$. In general, it is impossible to solve \eqref{modeeq} analytically, but we are able to introduce a formal adiabatic solution
\begin{equation}
    f_k(t)=\frac{1}{\sqrt{2\omega_k(t)}}\left(\alpha_k(t)e^{-\ri \int^t\omega_k(t')dt'}+\beta_k(t)e^{\ri\int^t\omega_k(t')dt'}\right).\label{adiabatic}
\end{equation}
 The auxiliary functions $\alpha_k(t),\beta_k(t)$ satisfy
\begin{align}
    \dot{\alpha}_k(t)=&\frac{\dot{\omega}_k}{2\omega_k}\beta_ke^{+2\ri \int^t \omega_k(t')dt'},\label{alphaeq}\\
    \dot{\beta}_k(t)=&\frac{\dot{\omega}_k}{2\omega_k}\alpha_ke^{-2\ri \int^t\omega_k(t')dt'},\label{betaeq}
\end{align}
and $|\alpha_k(t)|^2-|\beta_k(t)|^2=1$ for all $t$, which is equivalent to \eqref{KKnorm}. Note that the choice of the adiabatic solution is not unique, and we have chosen the zeroth order adiabatic solution with $V_k(t)=0$ in \cite{Habib:1999cs,Dabrowski:2014ica,Dabrowski:2016tsx}. We choose the initial condition $\alpha_k(t)\to 1$ and $\beta_k(t)\to 0$ as $t\to -\infty$ which realizes the adiabatic past vacuum.\footnote{The infinite past may be replaced by some finite $t$, which does not change as long as the initial time is sufficiently far from the time of the first particle production event.} In general, $\beta_k(t)$ becomes non-vanishing due to the time-dependence of background fields, which can be physically understood as ``particle production from vacuum''. We briefly review it in Appendix~\ref{PPreview}. The production of particles back-reacts to the dynamics of the background field. Throughout this work, we assume that the backreaction affects only the modulus $\phi$, and the background spacetime is intact.

Let us consider the dynamics of $\phi$ back-reacted by $\chi$-particles, whose E.O.M. is given by
\begin{equation}
    \ddot{\phi}(t)+3H\dot\phi+\lambda\langle\chi^2\rangle_{\rm ren}\phi(t)=0,
\end{equation}
where $\langle\hat{\chi}^2(x)\rangle_{\rm ren}$ is an expectation value of a renormalized $\hat{\chi}^2$ operator with the adiabatic vacuum state. The vacuum expectation value of $\hat\chi^2$ without renormalization is explicitly given by
\begin{align}
    \langle\chi^2\rangle=&\lim_{y\to x}\langle 0|_{\rm in} \hat{\chi}(x)\hat\chi(y)|0\rangle_{\rm in}\nonumber\\
    =&\lim_{y\to x}\int\frac{d^3k d^3k'}{(2\pi)^3a^3(t)}\delta^3({\bf k}-{\bf k}')\left(e^{\ri {\bf k}\cdot {\bf x}-\ri {\bf k}'\cdot{\bf y}}f_k(x^0)f_{k'}^*(y^0)\right)\nonumber\\
    =&\int\frac{d^3k}{(2\pi)^3a^3(t)}|f_k(t)|^2\nonumber\\
    =&\int\frac{d^3k}{(2\pi)^3a^3(t)}\frac{1}{2\omega_k(t)}\left(|\alpha_k(t)|^2+|\beta_k(t)|^2+\alpha_k(t)\beta_k^*e^{2\ri \int^t \omega_k(t')dt'}+\alpha_k^*(t)\beta_k e^{-2\ri \int^t \omega_k(t')dt'}\right)\nonumber\\
    =&\int\frac{d^3k}{(2\pi)^3a^3(t)}\frac{1}{2\omega_k(t)}\left(1+2|\beta_k(t)|^2+\alpha_k(t)\beta_k^*e^{2\ri \int^t \omega_k(t')dt'}+\alpha_k^*(t)\beta_k e^{-2\ri \int^t \omega_k(t')dt'}\right).
\end{align}
We note that $\beta_k(t)$ is vanishing unless particle production occurs, which cannot be removed by local counter terms. Therefore, we may identify the $\beta_k$-independent term to be a ``vacuum'' contribution. We first evaluate the ``vacuum'' contribution as
\begin{align}
    \int\frac{d^3k}{(2\pi)^3a^3(t)}\frac{1}{2\omega_k(t)}\to & \int \frac{d^dk}{(2\pi)^da^d(t)}\frac{\mu^{3-d}}{2\omega_k(t)}\nonumber\\
    =&\frac{2\pi^{\frac d 2}}{(2\pi)^d\Gamma\left(\frac d 2\right)}\int dk \frac{\mu^{3-d}k^{d-1}}{2\sqrt{k^2+M_{\rm eff}^2}}\nonumber\\
    =&\frac{M_{\rm eff}^2}{8\pi^2\epsilon}-\frac{M_{\rm eff}^2}{16\pi^2}\left[1-\gamma_E-\log\left(\frac{M_{\rm eff}^2}{4\pi\mu^2}\right)\right]+\mathcal{O}(\epsilon),
\end{align}
where we have used the dimensional regularization $d=3+\epsilon$ with $\epsilon\to0$, $M_{\rm eff}^2\equiv\lambda\phi^2-\frac{3}{2}\dot{H}-\frac{9}{4}H^2$ is the effective mass,\footnote{The effective mass $M_{\rm eff}^2$ may become tachyonic, which leads to imaginary part to the effective potential. Such instability simply shows the existence of tachyonic modes that cannot be naively integrated out. Therefore, one has to treat such modes separately if exist. In the following, we consider only the cases where such instability shows up.} and $\mu$ is renormalization scale parameter. We here applied dimensional regularization such that the invariance under trivial scale transformation $a\to ca$, ${\bm k}\to c{\bm k}$ ($\bm x\to c^{-1}\bm x$) with a constant $c$ holds. Thus, with appropriate counter terms, we can formally write the E.O.M. of $\phi$ as
\begin{equation}
   \ddot{\phi}(t)+\delta V(\phi)+F(t)=0,
\end{equation}
where $\delta V(\phi)$ denotes quantum corrected potential terms of $\phi$ and
\begin{align}
    F(t)\equiv& \lambda\phi(t)\int\frac{d^3k}{(2\pi)^3a^3(t)}\frac{1}{2\omega_k(t)}\left(2|\beta_k(t)|^2+\alpha_k(t)\beta_k^*e^{2\ri \int^t \omega_k(t')dt'}+\alpha_k^*(t)\beta_k e^{-2\ri \int^t \omega_k(t')dt'}\right)\nonumber\\
    \approx&\lambda\phi(t)\int\frac{d^3k}{(2\pi)^3a^3(t)}\frac{1}{\omega_k(t)}n_k(t),
\end{align}
where $n_k(t)\equiv |\beta_k(t)|^2$ and we have dropped fast-oscillatory terms since it would be averaged to be zero.\footnote{The neglected terms contain memory of the past, which makes the E.O.M. of $\phi$ an integro-differential equation.} We note that $|\beta_k(t)|$ typically decays faster than any powers of $k$, and therefore, there would be no UV divergence associated with it.\footnote{For adiabatic vacuum states $\beta(-\infty)=0$, $\beta_k(t)$ becomes non-zero due to Stokes phenomena/particle production, which is not a local but a global property with respect to time. Therefore, it is reasonable that the terms associated with such global (non-local) effects do not lead to UV divergences which are local effects due to short wave length modes.}

\begin{figure}[htbp]
    \centering
    \includegraphics[keepaspectratio, scale=0.8]{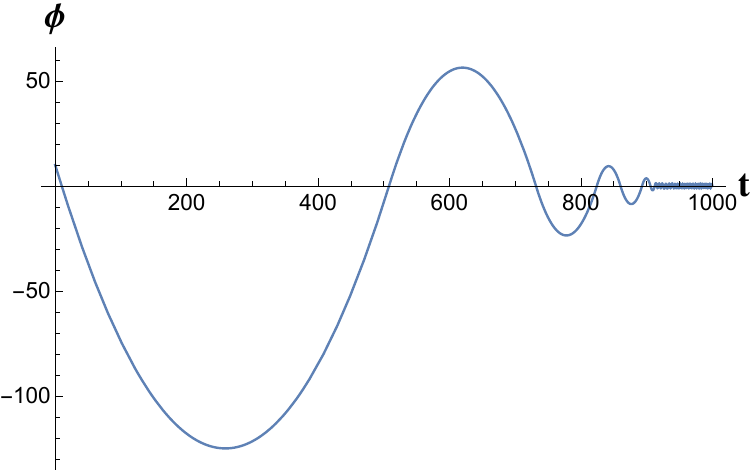}
    \caption{The moduli dynamics in flat spacetime. We have numerically solved the E.O.M and mode equations with the parameters $\lambda=1,\tilde{\phi}(0)=10,\tilde{\phi}'(0)=-1,N=2, n=10$. }
    \label{fig:example1}
    \end{figure}

We would like to give some comments about solving the dynamics of the system under consideration:
\begin{enumerate}
\item In the following, we assume that the quantum-corrected effective potential $\delta V(\phi)=0$ throughout this work. Such a situation is effectively realized e.g. if the theory is supersymmetric. Strictly speaking, the time-dependence of backgrounds break supersymmetry spontaneously and there would be some potential. We will leave the effect of the quantum corrected potential for future work.
    
    \item We use Emarkov-Milne equation, which directly yields the time-dependent particle number density: We use $f_k(t)=\xi_k(t) e^{-\ri \lambda_k(t)}$ as an Ansatz for the mode function. From the normalization condition~\eqref{KKnorm} the function $\lambda_k(t)$ satisfies
    \begin{equation}
        \lambda_k(t)=\frac12 \int^tdt'\xi_k^{-2}(t')
    \end{equation}
    and the mode equation reads
    \begin{equation}
    \ddot{\xi}_k(t)+\omega_k^2(t)\xi_k(t)=\frac14 \xi^{-3}_k(t),
    \end{equation}
     with the initial values
    \begin{equation}
        \xi_k(t_0)\to\frac{1}{\sqrt{2\omega_k(t)}},\quad \dot{\xi}_k(t_0)\to 0.
    \end{equation}
    In terms of $\xi_k(t)$, the particle number density $n_k(t)$ is simply given by
    \begin{equation}
        n_k(t)=\frac{\xi^2_k(t)}{2\omega_k(t)}\left(\left(\frac{1}{2\xi_k^2(t)}-\omega_k(t)\right)^2+\frac{\dot{\xi}_k^2(t)}{\xi_k^2(t)}\right),
    \end{equation}
    which is much easier to evaluate, since it can be solved simply as a set of differential equations. It is crucial to neglect the oscillatory terms in $F(t)$ to avoid an integro-differential equation.

      \item As done in \cite{Kofman:2004yc} we discretize the momentum integration appearing in $F(t)$, which allows us to perform numerical simulations. However, even with such approximation, it is still a hard problem unless we reduce the number of $k$-modes in numerical simulations. To do so, it would be useful to rescale the momentum $k$ by some reference scale. Assuming that the effective mass $M_{\rm eff}^2$ is dominated by the modulus coupling $\lambda\phi^2$, the most relevant scale turns out to be $v=\lambda^{\frac14}|\dot\phi(t_0)|^{\frac12}$ where $t_0$ is the time when $\phi$ first crosses $\phi=0$ at which $M_{\rm eff}\approx\lambda^{\frac12}|\phi|=0$.\footnote{We will explain the reason why $v$ is a reference scale below.} Therefore, we rescale dimensionful quantities by $v$ like $\tilde{k}=k/v$,
    \begin{equation}
\tilde{\omega}_k=\omega_k/v=\sqrt{\tilde{k}^2+\frac{M_{\rm eff}^2(t)}{v^2}}.
    \end{equation}
  Noting that
    \begin{equation}
    n_k(t)=\frac{\tilde{\omega}_k\tilde{\xi}^2_k}{2}\left(\left(\frac{1}{2\tilde{\omega}_k\tilde{\xi}_k^2}-1\right)^2+\frac{\dot{\tilde{\xi}}^2}{\tilde{\omega}^2_k\tilde{\xi}_k^2}\right),
    \end{equation}
    where $\tilde{\xi}_k=\sqrt{v}\xi_k$, we rewrite $F(t)$ as
    \begin{equation}
    F(t)=\lambda\phi(t)v^2\int\frac{d^3\tilde{k}}{(2\pi)^3}\frac{n_k(t)}{\tilde{\omega}_k(t)}=\frac{\lambda \phi(t)v^2}{2\pi^2}\int d\tilde{k}\frac{\tilde{k}^2n_k(t)}{\tilde{\omega}_k(t)}.
    \end{equation}
    Let us discuss how to perform the discretization of the integral effectively. We note that the particle number density after the first crossing of $m^2(t)=0$ is approximately given by
    \begin{equation}
        n_k=\exp\left(-\pi \frac{k^2}{\sqrt{\lambda}|\dot{\phi}(t_0)|}\right)=\exp\left(-\pi \tilde{k}^2\right).
    \end{equation}
    This is why we have taken $v$ as the reference scale.
     The mode $\pi\tilde k^2\sim \mathcal{O}(10)$ does not contribute to the integration and here we take $\tilde{k}=N$ to be the effective cut-off of the momentum integration where $N=\mathcal{O}(1)$ is a positive integer. Thus, $F(t)$ is approximately given by
    \begin{equation}
        F(t)\approx \frac{\lambda \phi(t)v^2}{2\pi^2}\int^{N}_0 d\tilde{k}\frac{\tilde{k}^2n_k(t)}{\tilde{\omega}_k(t)}\to\frac{\lambda \phi(t)v^2}{2\pi^2a^3(t)}\sum_{j=1}^{n}\left(\frac{N}{n}\right)^3\frac{j^2n_{k_j}(t)}{\tilde{\omega}_{k_j}(t)},
    \end{equation}
    where $n$ is a positive integer characterizing the lattice spacing and
    \begin{equation}
        \tilde{k}_j=\frac{Nj}{n}.
    \end{equation}
    We may recover the continuous case by taking $N,n\to\infty$. Thus, we have reduced the problem to be a set of $n$ differential equations.\footnote{We will not include the zero mode since it does not contribute to the effective potential within our approximation.}
\end{enumerate}
Using above approximation, we find a set of equations:
\begin{align}
&\tilde{\xi}_{k_i}''(\tilde{t})+\tilde{\omega}_{k_i}^2\tilde{\xi}_{k_i}(\tilde{t})=\frac{1}{4}\tilde{\xi}_{k_i}^{-3}(\tilde{t})\\
&\tilde{\phi}''(\tilde{t})+3\tilde{H}\tilde{\phi}(\tilde{t})+\frac{\lambda\tilde{\phi}(\tilde{t})}{2\pi^2a^3(\tilde{t})}\sum_{j=1}^n\left(\frac{N}{n}\right)^3\frac{j^2n_{k_j}(\tilde{t})}{\tilde{\omega}_{k_j}(\tilde{t})}=0,
\end{align}
where $i=1,2,\cdots, n$, $\tilde{t}=vt$, $\tilde{\phi}=\phi/v$, $\tilde{H}=H/v$ and the prime denotes derivative with respect to $\tilde{t}$. We show numerical solutions in Fig~\ref{fig:example1} with parameters and initial conditions $\lambda=1,\tilde{\phi}(0)=10,\tilde{\phi}'(0)=-1$ in Minkowski spacetime $a(t)=1$. As quoted in \cite{Kofman:2004yc}, in this model, parametric resonance occurs after the second and subsequent zero crossings, which extremely enhances the number density of low $k$ modes and strengthens the trapping effect. In numerical simulations, we have observed that the numerical solutions with different numbers of the mode number $n$ are qualitatively similar to each other but quantitatively different. 

\begin{figure}[htbp]
    \centering
    \includegraphics[keepaspectratio, scale=0.8]{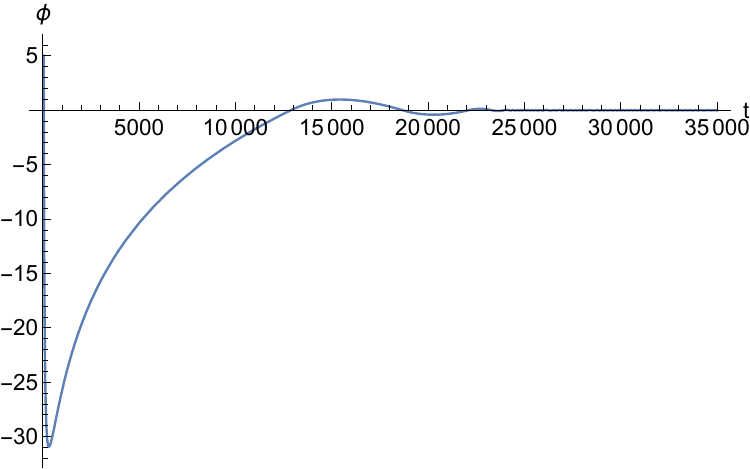}
    \caption{The moduli trapping effect in the matter dominated Universe. The parameters are chosen as $\tilde{t}_0=25$, $N=3$, $n=10$.}
    \label{fig:matter}
    \end{figure}

\begin{figure}[htbp]
    \centering
    \includegraphics[keepaspectratio, scale=0.8]{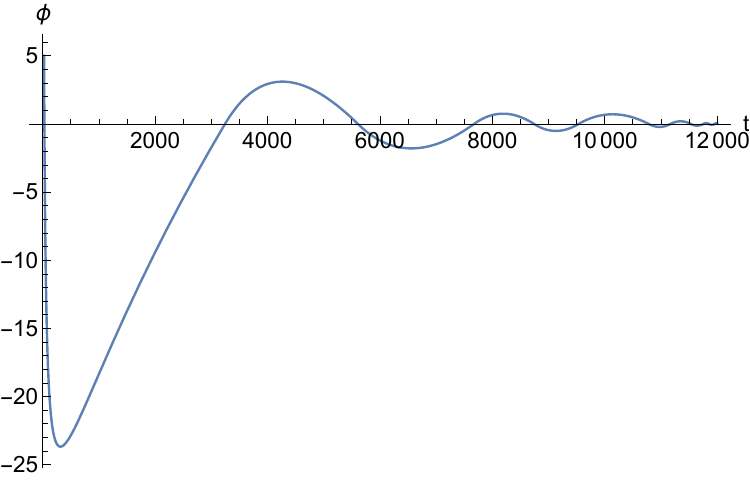}
    \caption{The moduli trapping effect in the radiation dominated Universe. The parameters are chosen as $\tilde{t}_0=55$, $N=3$, $n=9$.}
    \label{fig:radiation}
    \end{figure}

As another illustration, we consider the background
\begin{equation}
    a(t)=\left(\frac{t}{t_0}\right)^{\beta}=\left(\frac{\tilde{t}}{\tilde{t}_0}\right)^{\beta},
\end{equation}
where $\beta$ is a positive constant, and we have normalized the scale factor such that $a(\tilde{t}_0)=1$ at the initial time $\tilde{t}=\tilde{t}_0$. Then,
\begin{equation}
    \tilde{H}=\frac{\beta}{tv}=\frac{\beta}{\tilde{t}}.
\end{equation}
In Figs~\ref{fig:matter},\ref{fig:radiation} we show the numerical solution for the matter dominated universe $\beta=\frac23$ and the radiation dominated universe $\beta=1/2$, respectively. These figures show that the trapping mechanism works even in the expanding background. Note however that, if the initial Hubble parameter is large enough $(\tilde{t}_0\sim {\cal O}(1))$, it is possible to slow the trapping by the dilution of particles as well as Hubble friction. 
    
\begin{figure}[htbp]
    \centering
    \includegraphics[keepaspectratio, scale=0.8]{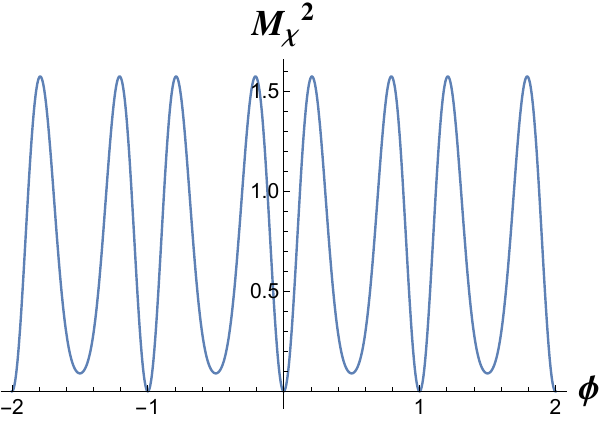}
    \caption{The behavior of the $\phi$-dependent mass~\eqref{toymass} for $\mu^2=1,A=0.7$. $\phi=n\in\mathbb{Z}$ are the true symmetric points and there are fake points, where the mass becomes small but non-zero.}
    \label{fig:mass}
    \end{figure}

    Let us consider a more involved case that instead of $\lambda\phi^2$ the mass term is given by
    \begin{equation}
M_\chi^2(\phi)=\mu^2(\sin\pi\phi+A\sin3\pi \phi)^2,\label{toymass}
    \end{equation}
    where $A$ is a real parameter. In this case, there is a true symmetric point $\phi=n\in \mathbb{Z}$ where $\chi$ becomes massless, but there is a fake symmetric point as shown in Fig.~\ref{fig:mass}. The E.O.M. of $\phi$ turns to 
\begin{equation}
\ddot{\phi}+2\pi\mu^2(\sin\pi\phi+A\sin3\pi\phi)(\cos\pi\phi+3A\cos3\pi\phi)\langle\chi^2\rangle=0.\label{multiEOM}
\end{equation}

\begin{figure}[htbp]
    \centering
    \includegraphics[keepaspectratio, scale=0.8]{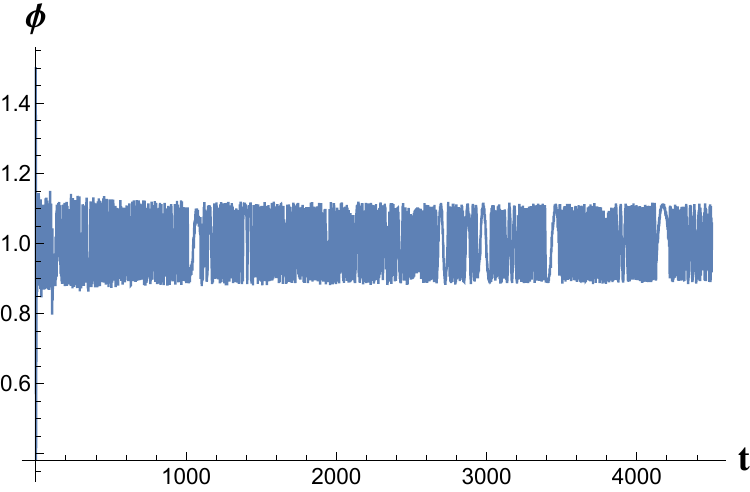}
    \caption{A numerical solution to~\eqref{multiEOM} with $\mu=10, n=10, A=0.5, \phi(0)=1.5,\dot\phi(0)=-1$.}
    \label{fig:multi}
    \end{figure}

\begin{figure}[htbp]
    \centering
    \includegraphics[keepaspectratio, scale=0.8]{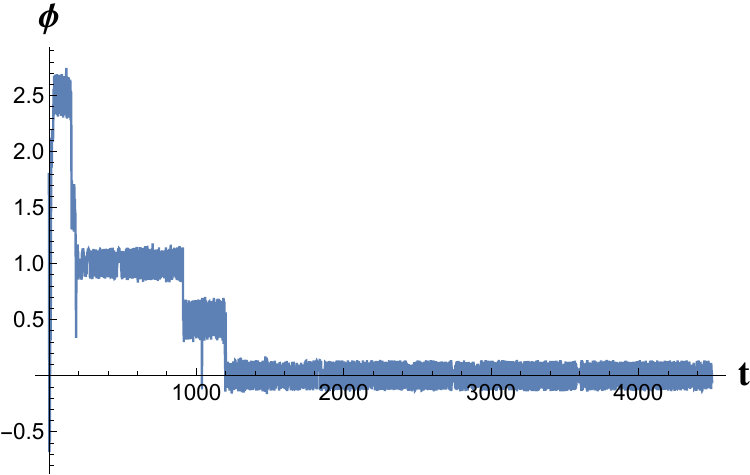}
    \caption{A numerical solution to~\eqref{multiEOM} with $\mu=7, n=10, A=0.5, \phi(0)=1.8,\dot\phi(0)=-1$.}
    \label{fig:multi2}
    \end{figure}

We show two numerical solutions with different parameters in Minkowski spacetime $a(t)=1$ in Figs~\ref{fig:multi}, \ref{fig:multi2}. We note that the false vacua are $\phi=\frac{2n+1}{2}$, ($n\in\mathbb{Z}$). In the former case, the modulus is immediately trapped to the true symmetric vacuum $\phi=1$, whereas the latter shows that the modulus is trapped both at the false and the true symmetric vacua, but finally reaches the true symmetric vacuum. We expect that modulus trapping at the true vacuum is not a generic property since the effective potential vanishes even at the false vacua, which can be seen from the E.O.M. of $\phi$. Nevertheless it is true that the particle production takes place more efficiently near the true vacua since $\chi$ becomes much lighter at the point. Therefore, if the true and false vacua are sufficiently separated, we expect the modulus to be trapped near the enhanced symmetry point.

Before closing this section, we finally comment on some technical issues remaining in our numerical approach shown here. We have tried our numerical simulations with different set of parameters or initial conditions several times. Although the behavior qualitatively converges, the results are quantitatively unstable under the change of parameters such as mode numbers. Such a behavior may be understood from the fact that parametric resonance become important when the modulus $\phi$ is almost trapped to the ESP, and the resonance is sensitive to the parameters such as momentum. As we will discuss later, the modular flavor symmetric models contain more issues in performing fully numerical simulations. Therefore, we will propose a semi-analytic approach to capture the moduli dynamics. Nevertheless, the numerical approach we have taken in this section would still be useful for some class of models.

\section{Modular flavor symmetric models}\label{Modularflavor}
    We show the modular flavor symmetric model considered in the following sections. In this section, we first briefly review modular flavor symmetry and then derive the E.O.M. of complex structure moduli as well as a spectator scalar $\chi$ while we will study in detail the $\chi$-particle production in the next section.
    
    \subsection{Modular flavor symmetry}

The $SL(2,Z)$ group,
\begin{align}
    \gamma=
    \begin{pmatrix}
    a & b \\
    c & d
    \end{pmatrix},
\end{align}
where $a,b,c,d$ are integer and $ad-bc=1$, 
is generated by two generators, $S$, and $T$,
\begin{align}
    S=
    \begin{pmatrix}
        0 & 1 \\ -1 & 0
    \end{pmatrix}, \qquad
    T=
    \begin{pmatrix}
     1 & 1 \\ 0 & 1   
    \end{pmatrix}.
\end{align}
They satisfy the following algebraic relations:
\begin{align}
    S^4=1, \qquad (ST)^3=1.
\end{align}
This group is referred to the homogeneous modular group, 
$\Gamma=SL(2,Z)$.

Under the modular symmetry, the modulus $\tau$ transforms as
\begin{align}
    \gamma(\tau) = \frac{a\tau + b}{c\tau + d}.
\end{align}
Note that $\tau$ is invariant under $S^2$.
That is, the generators satisfy
\begin{align}
    S^2=1, \qquad (ST)^3=1,
\end{align}
on the modulus $\tau$.
This group is referred to the inhomogeneous modular symmetry, $\bar \Gamma=PSL(2,Z)=SL(2,Z)/Z_2$.
Also we define the congruence subgroup $\Gamma(N)$,
\begin{align}
    \Gamma(N)=\left\{
    \begin{pmatrix}
        a & b \\ c & d
    \end{pmatrix}
    \in \Gamma ~\left|~
    \begin{pmatrix}
         a & b \\ c & d
    \end{pmatrix}
    =\begin{pmatrix}
        1 & 0\\ 0 & 1
    \end{pmatrix}
    \right.
    ({\rm mod}~N)
    \right\},
\end{align}
which includes $T^N$.
Similarly, we can define $\bar \Gamma(N)$.

The modular forms $f(\tau)_i$ are the holomorphic functions of $\tau$, 
which transform under the modular symmetry,
\begin{align}
    f(\gamma (\tau))_i=(c\tau +d)^k \rho_{ij}(\gamma)f(\tau)_j,
\end{align}
where $k$ is the modular weight and 
$\rho(\gamma)_{ij}$ is a unitary matrix.
Suppose that 
\begin{align}
    f(\gamma (\tau))_i=(c\tau +d)^k (\gamma)f(\tau)_i,
\end{align}
for $\gamma \in \bar \Gamma(N)$.
Then, the matrix $\rho(\gamma)_{ij}$ represents 
the quotient, $\Gamma_N=\bar \Gamma/\bar \Gamma(N)$, 
where $T^N=1$.
Interestingly, these quotients $\Gamma_N$ with $N=2,3,4,5$ are isomorphic to $S_3$, $A_4$, $S_4$, $A_5$, respectively. 

There are fixed points on $\tau$, i.e. 
$\tau =\ri, \omega=e^{2\pi \ri /3}, \ri \infty$, 
where residual symmetries remain.
That is, $Z_2^S, Z_3^{ST}, Z_N^T$ symmetries remain 
at $\tau =\ri, \omega=e^{2\pi \ri /3}, \ri \infty$, respectively.
The modular form $f(\tau)$ has definite charges under 
these residual symmetries and their behaviors are 
determined by their charges.

As illustrating examples, we study modular $A_4$ 
symmetric models in this paper.
Modular flavor symmetric superpotential can be written by 
\begin{align}
    W=Y_{ijk}(\tau)\Phi_i \Phi_j \Phi_k,
\end{align}
where $\Phi_i$ are the chiral superfields with modular weights $k_i$ and they have some representations under $A_4$.
The Yukawa couplings $Y_{ijk}(\tau)$ are modular forms.
The superpotential must be invariant under the 
modular symmetry including the $A_4$ symmetry.
The K\"ahler potential of chiral matter fields can be written by 
\begin{align}
    K=\sum_i\frac{1}{(2{\rm Im}\tau)^{k_i}}|\Phi_i|.
\end{align}

The fundamental modular forms of $A_4$ have the modular weight 2 and they correspond to the $A_4$ triplet $(Y_1(\tau),Y_2(\tau), Y_3(\tau))$ \cite{Feruglio:2017spp}.
Their explicit forms are shown in Appendix \ref{modular-form}.
The modular forms of higher modular weights can be obtained by their tensor products.
For example, we use the modular forms of weight 8, because all of three $A_4$ singlets, ${\bf 1}, {\bf 1}'$, and ${\bf 1}''$  
appear as modular forms when the weight is 8, 
i.e. $Y^{(8)}_{\bf 1}(\tau), Y^{(8)}_{{\bf 1}'}(\tau)$, and $Y^{(8)}_{{\bf 1}''}(\tau)$, which are shown in Appendix \ref{modular-form}.
The three $A_4$ singlets, ${\bf 1}, {\bf 1}'$, and ${\bf 1}''$  have the $Z_3^{ST}$ charges, 0, 1, 2, respectively.
The corresponding modular forms have additional contributions from the automorphic factors $(c\tau + d)^k$.

As a simple illustrating model, in the following sections we consider a single complex scalar spectator field that has the mass term
\begin{align}
|M_\chi(\tau,\bar{\tau})|^2|\chi|^2,
\end{align}
where $M_\chi(\tau,\bar\tau)|^2$ is given by modular forms. Such a complex scalar field can appear in supersymmetric models with
\begin{align}
    K=&\frac{1}{(2{\rm Im}\tau)^k}|\Phi|^2,\\
    W=&mY(\tau)\Phi S,
\end{align}
where $\Phi$ is a chiral superfield whose scalar component is $\chi$, $k$ is the modular weight of $\Phi$, $m$ a mass parameter, and $Y(\tau)$ a holomorphic function of $\tau$ given by modular forms. $S$ is an additional chiral superfield which is a singlet under modular symmetry.\footnote{Since $S$ is a singlet, it may have a heavy mass term independent of the above one. If $S$ has a heavy mass, it can decouple from the theory even at ESPs where $Y(\tau)\to0$.} Although there are fermions in such a model, which are as light as the scalar $\chi$, we only discuss the complex scalar $\chi$ only. The fermionic particle production can be similarly discussed along the line e.g. of \cite{Greene:2000ew,Peloso:2000hy,Adshead:2015kza} but would be more involved than the bosonic case we consider here because of the chiral structure. 

We would like to comment on the relation to the standard model. In the supersymmetric standard model, mass terms of scalars and fermions through Yukawa couplings are not generated until the electroweak symmetry is spontaneously broken. Therefore, we do not expect standard model particles to contribute the moduli trapping.\footnote{Scalars would have mass terms by supersymmetry breaking and they generally depend on moduli. Therefore, the moduli dynamics may generate scalars in the supersymmetric standard model.} Instead, we can identify $\Phi$ as a superfield that obtains its mass term through the GUT symmetry breaking\footnote{Another candidate would be right-handed (s)neutrinos, where the superpotential can be written by $W=mY(\tau)\Phi^2$.}. Therefore, the mass parameter $m$ can be sufficiently large (but smaller than $M_{\rm pl}$).

    \subsection{Dynamics of moduli and spectator scalars}
    We consider quantization of a spectator complex scalar field $\chi$ that couples to the complex structure modulus $\tau$, and the action of $\chi$ is given by
\begin{equation}
    S=-\int d^4x\sqrt{-g}\left[\frac{1}{(2{\rm Im}\tau)^k}|\partial\chi|^2+|M_\chi(\tau,\bar\tau)|^2|\chi|^2\right].
\end{equation}
We introduce quasi-canonical real fields
    \begin{equation}
        \tau=\sqrt{2}\theta/M_{\rm pl}+\ri e^{\sqrt2 \phi/M_{\rm pl}},
    \end{equation}
    which will become a choice that makes $\phi$ canonical. We assume that the ``bare mass'' $M^2(\tau,\bar\tau)$ is given by modular forms as
\begin{equation}
    |M_\chi(\tau,\bar\tau)|^2=m^2\left|Y^{(8)}_{\bm 1}(\tau)\right|^2 \ {\rm or} \ m^2\left|Y^{(8)}_{\bm 1'}(\tau)\right|^2,
\end{equation}
where $m$ denotes some mass scale such as the VEV of a  GUT Higgs field, and the background spacetime is $ds^2=-dt^2+a^2(t)d{\bm x}^2$. Introducing $\chi=\tilde{\chi}_1+\ri \tilde{\chi}_2$, we rewrite the action as
\begin{align}
    S=&\int dtd^3x\left[\frac{1}{2^k}a^3e^{-\sqrt{2}k\phi/M_{\rm pl}}\left(\dot{\tilde{\chi}}_i^2-\frac{1}{a^2}({\bm \nabla}\tilde{\chi}_i)^2\right)-a^3 |M_\chi(\tau,\bar\tau)|^2\tilde{\chi}^2_i\right],
\end{align}
which can be made canonical by introducing a new basis $\tilde{\chi}_i=\alpha\chi_i$ where 
\begin{equation}
    \alpha\equiv \frac{2^{\frac{k-1}{2}}}{a^{\frac32}}e^{\frac{\sqrt2}{2}k\phi/M_{\rm pl}}.
\end{equation}
With the new basis, we find
\begin{align}
    S=\sum_{i=1}^2\int dtd^3x\left[\frac{1}{2}\left(\dot\chi_i^2-\frac{1}{a^2}({\bm \nabla}\chi_i)^2\right)-\frac{1}{2}M_{\rm eff}^2\chi_i^2\right],
\end{align}
where we have assumed that the background field depends only on $t$, and used integration by parts in the second equality. Thus, the effective mass of the canonically normalized spectator field is
\begin{align}
    M_{\rm eff}^2=&2^ke^{\sqrt2 k\phi/M_{\rm pl}} |M_\chi|^2-2\frac{\dot\alpha^2}{\alpha^2}+\frac{\ddot\alpha}{\alpha}\nonumber\\
    =&2^ke^{\sqrt2 k\phi/M_{\rm pl}} |M_\chi|^2-\frac{k^2}{2M_{\rm pl}^2}\dot\phi^2+\frac{3\sqrt2 k}{2M_{\rm pl}}\dot\phi H+\frac{\sqrt2 k}{2M_{\rm pl}}\ddot\phi-\frac{9}{4}H^2-\frac32 \dot H\nonumber\\
    \approx&2^ke^{\sqrt2 k\phi/M_{\rm pl}} |M_\chi|^2.
\end{align}
The time derivative terms of $\phi$ in the second line may be interpreted as the higher-dimensional curvature induced mass as ${\rm Im}\tau$ is related to the area of the extra dimensional torus. We will neglect such terms as well as Hubble induced terms in the following.\footnote{Since the mass becomes zero at the critical point and the low $k$-mode can be tachyonic, it is generally non-trivial if such approximation is allowed. We may neglect such effect by assuming that the spectator scalar has a small mass that is comparable to the derivative terms so that the tachyonic instability disappears. On the other hand, the instability near the critical point may enhance the particle production.} We also notice that the effective mass is modular invariant as it should be.

Let us consider the effective action of the complex structure modulus $\tau$ given by
    \begin{align}
        S&=-\int d^4x\sqrt{-g} \left[\frac{M_{\rm pl}^2}{(2{\rm Im}\tau)^2}|\partial\tau|^2+V(\tau,\bar\tau)+a^{-3}M_{\rm eff}^2\sum_{i=1}^2\langle\chi_i^2\rangle\right]\nonumber\\
        &=-\int d^4x \sqrt{-g}\left[\frac12 (\partial\phi)^2+\frac12 e^{-2\sqrt{2}\phi/M_{\rm pl}}(\partial\theta)^2+V(\tau,\bar\tau)+a^{-3}M_{\rm eff}^2\sum_{i=1}^2\langle\chi_i^2\rangle\right].
    \end{align}
    Note that the appearance of non-covariant expression $+a^{-3}M_{\rm eff}^2\sum_{i=1}^2\langle\chi_i^2\rangle$ is the result of making $\chi_i$ canonical.\footnote{One could use the original variable, but would be more involved since both the mass and the kinetic term of $\chi$ appear in the E.O.M. of $\phi$ through the non-minimal coupling in the kinetic term.}
    The equations of motions of $\phi$, $\theta$ are given by
\begin{align}
&-\Box \phi-\sqrt{2}e^{-2\sqrt2 \phi/M_{\rm pl}}(\partial\theta)^2-\frac{2\sqrt{2}e^{\sqrt2\phi/M_{\rm pl}}}{M_{\rm pl}}{\rm Im}(\partial_\tau V)+\frac{1}{a^3}\partial_\phi M_{\rm eff}^2\sum_{i=1}^2\langle\chi_i^2\rangle=0,\\
&-\Box\theta+\frac{2\sqrt2 }{M_{\rm pl}}\partial_\mu\phi\partial^\mu\theta+\frac{2\sqrt{2}e^{2\sqrt2\phi/M_{\rm pl}}}{M_{\rm pl}}{\rm Re}(\partial_\tau V)+e^{2\sqrt2 \phi/M_{\rm pl}}\frac{1}{a^3}\partial_\theta M_{\rm eff}^2\sum_{i=1}^2\langle\chi_i^2\rangle=0.
\end{align}

It is straightforward to generalize our previous discussion to this case by using
\begin{equation}
    \omega_p^2=\frac{p^2}{a^2}+M_{\rm eff}^2\approx\frac{p^2}{a^2}+2^ke^{\sqrt2 k\phi/M_{\rm pl}} |M_\chi|^2 .
\end{equation}
The vacuum expectation value of $\langle\hat{\chi}_1^2\rangle$ is given by
\begin{align}
    \langle\hat{\chi}_1^2\rangle\approx \frac{a^3M_{\rm eff}^2}{8\pi^2\epsilon}-\frac{a^3M_{\rm eff}^2}{16\pi^2}\left[1-\gamma_E-\log\left(\frac{M_{\rm eff}^2}{4\pi\mu^2}\right)\right]+\int\frac{d^3p}{(2\pi)^3}\frac{n_p(t)}{\omega_p(t)}.
\end{align}
Thus, the renormalized effective potential is found to be
\begin{align}
    V_{\rm eff}=&\frac{M_{\rm eff}^4}{8\pi^2\epsilon}-\frac{M_{\rm eff}^4}{16\pi^2}\left[1-\gamma_E-\log\left(\frac{M_{\rm eff}^2}{4\pi\mu^2}\right)\right]+\delta_{\rm CT}+\frac{M_{\rm eff}^2}{a^3}\int\frac{d^3p}{(2\pi)^3}\frac{n_p(t)}{\omega_p(t)},\label{modulareff}
\end{align}
where $\delta_{\rm CT}$ denotes possible local counter-terms eliminating divergent pieces.\footnote{We point out that the ``effective potential'' contains the derivative of the background field $\phi$ in general, which behaves as higher-derivative terms. Such terms should be removed by appropriate counter-terms in order to avoid (possibly) unphysical ghost degrees of freedom.} As the previous sections, we assume the cancellation of all but the last term in \eqref{modulareff}.

In analysing the dynamics of the moduli fields $\phi,\theta$, there are several technical issues in this model in addition to the ones discussed within toy models: Since the coupling between $\chi$ and moduli are suppressed by Planck scale which causes hierarchically small or large numbers, and the moduli dependence is quite complicated, the numerical costs become more than that in the toy models. More specifically, the effective field theory description requires the field velocity and the Hubble parameter to be much smaller than the Planck scale. Furthermore, if the moduli trapping occurs after inflation, the Hubble scale needs to be less than about ${\cal O}(10^{13}{\rm GeV})$ by the constraint on the tensor-to-scalar ratio. Therefore, it would be natural to assume the ratio between the initial field velocity and Planck scale to be below $\mathcal{O}(10^{-5})$, which appear in couplings if we normalize all dimensionful parameters by the scale of initial field velocity. The complication of the mass term given by modular forms shown later further makes the numerical simulations difficult. Therefore, we will develop semi-analytic approaches to discuss the moduli dynamics within modular flavor symmetric models in the next section.

\section{Particle production at ESPs in modular flavor symmetric models}\label{analyticPP}
In order to overcome some technical difficulties within our setup, we consider a semi-analytic approach, where we analytically estimate the number density of the spectator scalar particle produced at the first particle production time at which $M^2_\chi(t_0)\approx 0$. Using the estimate, we analytically evaluate the effective potential arising after the first particle production event, and numerically solve the E.O.M. of moduli fields with the estimated effective potential, which does not contain the difficulties mentioned before. Although this approach cannot capture the subsequent particle production events, we expect such events just strengthen the trapping potential, which does not change the dynamics of moduli qualitatively.

We first show analytic formulas for particle number density produced at the crossing of the ESP. To do so, we need to know the behavior of the effective mass near the ESPs. In modular symmetric models, the ESPs are $\tau=\ri,e^{2\ri \pi/3}(\equiv\omega),+\ri \infty$ where some modular forms vanish. In the following discussion, we focus on $\tau=\omega$ as a representative case, but similar analysis can be done for any other critical points in a similar way.

In the following subsections, we first discuss the behavior of the effective mass of the spectator scalar field near the ESP~$\tau\sim\omega$, which enables us to analytically estimate the amount of particle number density produced after the crossing with the ESP.

With the aid of the analytic expression of the effective mass around $\tau\sim\omega$, we discuss one-dimensional dynamics of $\theta$ or $\phi$ where one of them is fixed to a constant value. We give analytic formulas for the particle number density as well as the effective forces of the produced particle on the modulus.

\subsection{The behavior of the effective mass near the critical point}
The modular forms can be classified as representations of subgroup. The singlet modular forms of weight 8, $Y^{(8)}_{{\bm 1},{\bm 1'},{\bm 1''}}$ transform as \cite{Novichkov:2021evw}
\begin{align}
    \left(\begin{array}{c} Y^{(8)}_{\bm 1}(-(\tau+1)^{-1})\\ Y^{(8)}_{{\bm 1'}}(-(\tau+1)^{-1}) \\  Y^{(8)}_{{\bm 1''}}(-(\tau+1)^{-1})
    \end{array}\right)=(-(\tau+1))^{8}\left(\begin{array}{ccc}
        1 & 0 &0  \\
         0 & \omega&0\\
         0&0&\omega^{2}
    \end{array}\right) \left(\begin{array}{c}Y^{(8)}_{{\bm 1}}(\tau) \\ Y^{(8)}_{{\bm 1'}}(\tau)\\ Y^{(8)}_{{\bm 1''}}(\tau)
    \end{array}\right),
\end{align}
which can be equivalently written as
\begin{align}
    \left(\begin{array}{c} Y^{(8)}_{\bm 1}(\omega^2u)\\ Y^{(8)}_{{\bm 1'}}(\omega^2u)\\  Y^{(8)}_{{\bm 1''}}(\omega^2u)
    \end{array}\right)=\left(\frac{1-\omega^2 u}{1-u}\right)^8\left(\begin{array}{ccc}
        \omega^{-8} & 0 &0  \\
         0 & \omega^{-7} &0\\
         0&0&\omega^{-6}
    \end{array}\right) \left(\begin{array}{c}Y^{(8)}_{{\bm 1}}(u) \\ Y^{(8)}_{{\bm 1'}}(u)\\ Y^{(8)}_{{\bm 1''}}(u)
    \end{array}\right),
\end{align}
where 
\begin{equation}
    u\equiv \frac{\tau-\omega}{\tau-\omega^2},
\end{equation}
is a variable for the deviation from the symmetric point $\tau=\omega$. Therefore, we obtain
\begin{align}
    \left(\begin{array}{c} \tilde{Y}^{(8)}_{\bm 1}(\omega^2u)\\ \tilde{Y}^{(8)}_{{\bm 1'}}(\omega^2u)\\  \tilde{Y}^{(8)}_{{\bm 1''}}(\omega^2u)
    \end{array}\right)=\left(\begin{array}{ccc}
        \omega^{-8} & 0 &0  \\
         0 & \omega^{-7} &0\\
         0&0&\omega^{-6}
    \end{array}\right) \left(\begin{array}{c}\tilde{Y}^{(8)}_{{\bm 1}}(u) \\ \tilde{Y}^{(8)}_{{\bm 1'}}(u)\\ \tilde{Y}^{(8)}_{{\bm 1''}}(u)
    \end{array}\right),
\end{align}
where $\tilde{Y}^{(8)}_{\bm r}(u)\equiv (1-u)^{-8}Y^{(8)}_{\bm r}(u)$. Expanding both sides with respect to $\omega$ yields
\begin{equation}
   \left( \omega^{2l}-\omega^{q_{\bm r}-8}\right)\frac{d^l\tilde{Y}^{(8)}_{\bm r}(u)}{du^l}\biggr|_{u\to0}=0,
\end{equation}
where $q_{\bm r}=0,1,2$ for ${\bm 1},{\bm 1'}, {\bm 1''}$, respectively. This relation implies that  $\frac{d^l\tilde{Y}^{(8)}_{\bm r}(u)}{du^l}\biggr|_{u\to0}=0$ unless $2l=q_{\bm r}-8 \ (\text{mod 3})$. Thus, $\tilde{Y}^{(8)}_{\bm 1,\bm 1'}(0)=0$, and $\frac{d\tilde{Y}^{(8)}_{\bm 1,\bm 1''}}{du}(0)=0$. Noting that 
\begin{align}
    \frac{\partial u}{\partial \tau}=\frac{(1-u)^2}{\sqrt3 \ri},
\end{align}
we find
\begin{align}
   & Y^{(8)}_{\bm r}(\tau)|_{\tau\to\omega}=\tilde{Y}^{(8)}_{\bm r}(u)|_{u\to 0},\\
   & \frac{dY^{(8)}_{\bm r}}{d\tau}\Biggr |_{\tau\to \omega}=\frac{1}{\sqrt3 \ri}\left[-8\tilde{Y}^{(8)}_{\bm r}+\frac{d\tilde{Y}^{(8)}_{\bm r}}{du}\right]_{u\to0},\\
   &\frac{d^2Y^{(8)}_{\bm r}}{d\tau^2}\Biggr|_{\tau\to \omega}=\left[-24\tilde{Y}^{(8)}_{\bm r}+6\frac{d\tilde{Y}_{\bm r}^{(8)}}{du}-\frac13\frac{d^2\tilde{Y}_{\bm r}^{(8)}}{du^2}\right]_{u\to0}.
\end{align}
Note also that
\begin{align}
    u=\frac{\tau-\omega}{\omega-\omega^2}+\mathcal{O}((\tau-\omega)^2)=\frac{\tau-\omega}{\sqrt3 \ri}+\mathcal{O}((\tau-\omega)^2).
\end{align}
The leading order terms of each singlet ${\bm 1},{\bm 1'}, {\bm 1''}$ are as follows:
\begin{align}
    Y^{(8)}_{\bm 1}(\tau)=&-\frac16\frac{d^2\tilde{Y}_{\bm 1}^{(8)}(0)}{du^2} (\tau-\omega)^2+\mathcal{O}((\tau-\omega)^3),\\
     Y^{(8)}_{\bm {1'}}(\tau)=&\frac{1}{\sqrt3 \ri}\frac{d\tilde{Y}^{(8)}_{\bm 1'}(0)}{du}(\tau-\omega)+\mathcal{O}((\tau-\omega)^2),\\
     Y^{(8)}_{\bm {1''}}(\tau)=&\tilde{Y}^{(8)}_{\bm {1''}}(0)+\mathcal{O}((\tau-\omega)^1),
\end{align}
and the last one shows that $\bm 1''$ cannot produce the particles at ESP $\tau=\omega$, because $\chi$ is still massive.\footnote{More precisely speaking, for $\bm r = \bm 1 ''$, particles are produced if $\tau=\omega$ is the local minimum of $Y_{\bm 1''}^{(8)}$, but as the leading term is non vanishing, the particle production would be less than that of $\bm r= \bm 1 \ \text{or}\ \bm 1'$}.
Thus, the effective mass can be approximated as
\begin{align}
    M_{\rm eff}^2=&\frac{2^6|C|^2}{9} m^2 e^{8\sqrt2\phi/M_{\rm pl}}|\tau-\omega|^4+\cdots\nonumber\\
 =& \frac{9|C|^2}{4} m^2 \left[\left(\frac{\sqrt2\theta}{M_{\rm pl}}+\frac12\right)^2+\left(e^{\sqrt2 \phi/M_{\rm pl}}-\frac{\sqrt3}{2}\right)^2\right]^2+\cdots  (\text{for ${\bm r = \bm 1}$}),\\
    M_{\rm eff}^2=&\frac{2^8|D|^2}{3} m^2e^{8\sqrt2 \phi/M_{\rm pl}} |\tau-\omega|^2+\cdots\nonumber\\
  =&27|D|^2 m^2 \left[\left(\frac{\sqrt2\theta}{M_{\rm pl}}+\frac12\right)^2+\left(e^{\sqrt2 \phi/M_{\rm pl}}-\frac{\sqrt3}{2}\right)^2\right]+\cdots (\text{for ${\bm r = \bm 1'}$}),
\end{align}
where $C\equiv\frac{d^2\tilde{Y}^{(8)}_{\bm 1}(0)}{du^2}$, $D\equiv \frac{d\tilde{Y}^{(8)}_{\bm 1'}(0)}{du}$, and ellipses denote the terms higher order in $(\tau-\omega)$.

Before going to the details of particle production, we would like give a few comments on the behavior of the effective mass for each representation. We notice that depending on the representation $\bm r$, the behavior of the mass term near the ESP $\tau=\omega$ changes. Since the moduli dependence comes with powers of $\chi/M_{\rm pl}$ or $\theta/M_{\rm pl}$, we can say the particle production in $\bm r=\bm 1$ case should be much smaller than that in $\bm r =\bm 1 '$. Therefore, the moduli trapping effect becomes more significant for $\bm r ={\bm 1}'$. Such observation is based on the fact that moduli are gravitationally coupled to matter fields, which are completely independent of particle production dynamics discussed below. 

\subsection{1D dynamics: $\phi$-fixed}
In this subsection, we discuss the particle production due to the $\theta$ dynamics while $\phi$ is fixed at the ESP $\phi=\phi_0$ where $e^{\sqrt2 \phi_0/M_{\rm pl}}=\frac{\sqrt3}{2}$, and we parametrize $\theta$ as
\begin{equation}
    \theta=v(t-t_0)-\frac{1}{2\sqrt2}M_{\rm pl},
\end{equation}
where $t_0$ is the time at which $\theta$ crosses the enhanced symmetry point and $v>0$ is the velocity of $\theta$ at $t=t_0$. This parametrization is a good approximation if Hubble parameter is smaller than $\sqrt{v}$. (See appendix~\ref{freedynamics}.) Then, near $t\sim t_0$ the effective mass can be approximated by
\begin{align}
    M_{\rm eff}^2\approx & 9|C|^2m^2 \left(\frac{v(t-t_0)}{M_{\rm pl}}\right)^4\ (\text{for ${\bm r = \bm 1}$}),\label{1Dtheta1}\\
    M_{\rm eff}^2\approx & 54|D|^2 m^2 \left(\frac{v(t-t_0)}{M_{\rm pl}}\right)^2\ (\text{for ${\bm r = \bm 1'}$}),\label{1Dtheta1p}
\end{align}
and the effective frequency for ${\bm k}$-mode is 
\begin{equation}
    \omega_{k}^2\approx \frac{k^2}{a^2(t_0)}+M_{\rm eff}^2,
\end{equation}
where we have approximated the scale factor by its value at $t=t_0$. We will show the analytic estimate of the particle production taking place around $t\sim t_0$ separately for both $\bm r =\bm 1$ and $\bm r = \bm 1'$ respectively.

\vspace{10pt}
\fbox{$\bm r =\bm 1$ case}

\begin{figure}[htbp]
    \centering
    \includegraphics[keepaspectratio, scale=0.8]{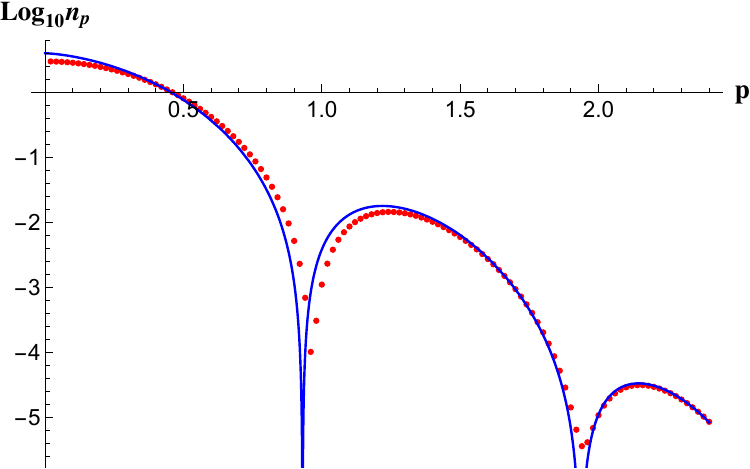}
    \caption{Comparison between the analytic formula~\eqref{theta1n} (the blue line) and  numerical results (red dots). The momentum $p$ is in units of the velocity $\sqrt{|v|}$ at the particle production time $t_0$. In this example, we have taken $vt_0=50$, $vt_{\rm ini}=15$, $\beta=\frac12$, $\sqrt{|v|}/M_{\rm pl}=5\times10^{-4}$, $m/\sqrt{|v|}=5\times10^2$. We have also checked that the agreement is quite well for various sets of parameters.} 
    \label{fig:theta1comp}
    \end{figure}

In order to estimate the particle production rate, we need to find the turning points at which $\omega_k^2(t)=0$ in a complex $t$-plane. (See~\cite{Dumlu:2010ua,Dabrowski:2014ica,Dabrowski:2016tsx,Enomoto:2020xlf,Taya:2020dco,Hashiba:2021npn,Yamada:2021kqw} for review.) With \eqref{1Dtheta1} the turning points are found to be
\begin{equation}
    t_1^{\pm}-t_0=e^{\pm\ri \pi/4} \frac{M_{\rm pl}}{v} \left(\frac{k^2}{9|C|^2m^2a^2(t_0)}\right)^{\frac14},\quad t_2^{\pm}-t_0=e^{\pm3\ri \pi/4} \frac{M_{\rm pl}}{v} \left(\frac{k^2}{9|C|^2m^2a^2(t_0)}\right)^{\frac14},
\end{equation}
with which the effective frequency can be rewritten as
\begin{align}
    \omega_k^2\approx R\left(-(t_*-t_0)^4+(t-t_0)^4\right),
\end{align}
where $t_*$ can be any of turning points and
\begin{equation}
    R\equiv \frac{9|C|^2m^2v^4}{M_{\rm pl}^4}.
\end{equation}
Notice that there are two pairs of turning points $(t_i^+,t_i^-)$ $(i=1,2)$ which are complex conjugate to each other, and Stokes lines connecting them crosses the real $t$-axis. The amount of particle production can be approximately given by a simple formula (see e.g.~\cite{Dumlu:2010ua,Dabrowski:2014ica,Dabrowski:2016tsx,Enomoto:2020xlf,Taya:2020dco,Hashiba:2021npn,Yamada:2021kqw} for details\footnote{See also~\cite{Froman:1970toy}.})
\begin{equation}
   n_k(t)= |\beta_k(t_i)|^2\approx \left|e^{F_1}+e^{\ri \theta_{12}}e^{F_2}\right|^2,
\end{equation}
where we have assumed that $t$ is sufficiently later than $t_0$, and
\begin{align}
    F_i\equiv&\exp\left(\ri \int_{t_i^-}^{t_i^+}\omega_k(t')dt'\right),\\
\theta_{12}\equiv&\exp\left(2\ri \int_{t_2^+}^{t_1^+}\omega_k(t')dt'\right).
\end{align}
More explicitly we find
\begin{align}
   F_i=&\exp\left[-\frac{\Gamma^2(\frac14)}{6\sqrt\pi R^{\frac14}}\left(\frac{k}{a(t_0)}\right)^{\frac32}\right],\\
\theta_{12}=&\frac{\Gamma^2(\frac14)}{3\sqrt\pi R^{\frac14}}\left(\frac{k}{a(t_0)}\right)^{\frac32}.
\end{align}
Therefore, the particle number density after crossing two Stokes lines can be approximately estimated as
\begin{align}
    \Delta n_k=2e^{-\gamma_k}(1+\cos\gamma_k),\label{theta1n}
\end{align}
where 
\begin{equation}
\gamma_k\equiv \frac{\Gamma^2(\frac14)}{3\sqrt\pi R^{\frac14}}\left(\frac{k}{a(t_0)}\right)^{\frac32}.
\end{equation}
We have numerically checked the validity of this formula, and find a very good agreement of the numerical result and the analytic formula as shown in Fig.~\ref{fig:theta1comp}. Thus, we can estimate the total particle production as well as the effective potential on the basis of the approximate formula~\eqref{theta1n}.

With the approximate particle number formula for a $k$-mode~\eqref{theta1n}, we are able to estimate the total number density to be\footnote{The oscillatory part $e^{\gamma_k}\cos\gamma_k$ turns out to be vanishing in the momentum integration.}
\begin{align}
    \Delta N=&\int \frac{d^3k}{(2\pi)^3}\Delta n_k=\frac{2\times 4\pi}{8\pi^3}\left(\frac{\Gamma^2(\frac14)}{3\sqrt\pi R^{\frac14} a^{\frac32}(t_0)}\right)^{-2}\times\frac{2}{3}=\frac{6\sqrt R a^3(t_0)}{\pi \Gamma^4(\frac14)}.
\end{align}
Even though we have an approximate formula of the produce particle number density, it is still difficult to evaluate the effective force given by
\begin{equation}
    F_\theta(t)=\frac{{(\rm Im}\tau)^2\partial_\theta (M_{\rm eff}^2)}{a^3}\int\frac{d^3p}{(2\pi)^3}\frac{\Delta n_p}{\omega_p(t)}.
\end{equation}
Since the effective support of the integrand is localized to small $k$, we may approximate $\omega_p\approx M_{\rm eff}$ in the integrand, which allows us to approximate $F_\theta(t)$ as
\begin{equation}
    F_\theta(t)\approx\frac{3\partial_\theta (M_{\rm eff}^2)}{4a^3M_{\rm eff}}\Delta N\Theta(t-t_0)=\frac{9\sqrt R a^3(t_0)\partial_\theta (M_{\rm eff}^2)}{2\pi \Gamma^4(\frac14)M_{\rm eff}a^3(t)}\Theta(t-t_0),
\end{equation}
where we have introduced the Heaviside theta function $\Theta(t-t_0)$.

\vspace{10pt}
\fbox{$\bm r =\bm 1'$ case}

\begin{figure}[htbp]
    \centering
    \includegraphics[keepaspectratio, scale=0.8]{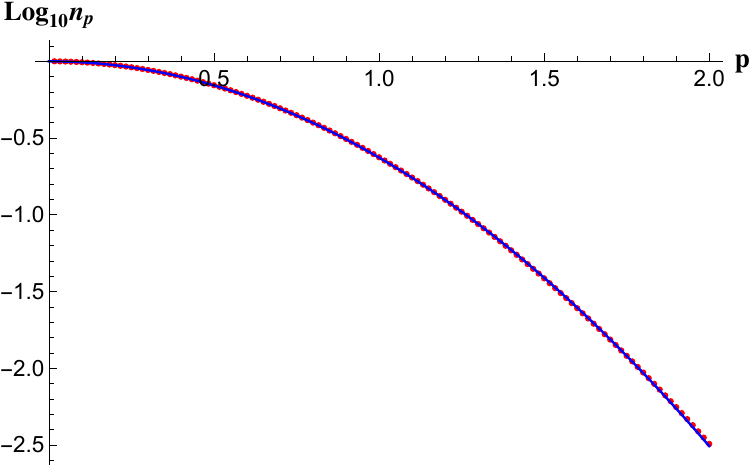}
    \caption{Comparison between the analytic formula~\eqref{theta1pn} (the blue line) and numerical results (red dots). The momentum $p$ is in units of the velocity $\sqrt{|v|}$ at the particle production time $t_0$. Here, we have taken $vt_0=70$, $vt_{\rm ini}=5$, $\beta=\frac12$, $\sqrt{|v|}/M_{\rm pl}=10^{-4}$, $m/\sqrt{|v|}=10$. We have also checked that the agreement is quite well for various sets of parameters.} 
    \label{fig:theta1pcomp}
    \end{figure}

Next, we consider the case $\bm r =\bm 1'$ where the effective mass is given by~\eqref{1Dtheta1p}, and find a pair of turning points
\begin{align}
    t^\pm_1-t_0=\pm\ri\frac{M_{\rm pl}k}{3\sqrt6 |D|m v}.
\end{align}
Similarly to the previous case, the particle number produced at $t=t_0$ can be evaluated by
\begin{align}
    n_k(t_0)\approx& \exp\left(2\ri \int_{t_1^-}^{t_1^+}\omega_k(t')dt'\right)\nonumber\\
    =&\exp \left[-\frac{\pi k^2M_{\rm pl}}{3\sqrt6 |D|mv}\right].\label{theta1pn}
\end{align}
We show the comparison of our analytic formula~\eqref{theta1pn} and numerical results in Fig.~\ref{fig:theta1pcomp}. Again, we have found an excellent agreement between them.

Accordingly, the total particle density is found to be
\begin{equation}
    \Delta N=\int \frac{d^3k}{(2\pi)^3}n_k(t_0)=\frac{1}{2\pi^2}\times\frac{\sqrt\pi}{4}\left(\frac{3\sqrt6|D|mv}{\pi M_{\rm pl}}\right)^{\frac32}=\frac{1}{8\pi^3}\left(\frac{3\sqrt6|D|mv}{M_{\rm pl}}\right)^{\frac32},
\end{equation}
and the effective force is approximately given by
\begin{align}
    F_\theta(t)=\frac{3a^3(t_0)\partial_\theta (M_{\rm eff}^2)}{4a^3(t)}\int\frac{d^3p}{(2\pi)^3}\frac{n_p(t)}{\omega_p(t)}\approx \frac{3a^3(t_0)(\partial_\theta M_{\rm eff}^2)(t)}{32\pi^3 a^3(t)M_{\rm eff}(t)}\left(\frac{3\sqrt6|D|mv}{M_{\rm pl}}\right)^{\frac32}\Theta(t-t_0).
\end{align}




\subsection{1D dynamics: $\theta$-fixed}
\begin{figure}[htbp]
  \begin{minipage}[b]{0.55\linewidth}
    \centering
    \includegraphics[keepaspectratio, scale=0.5]{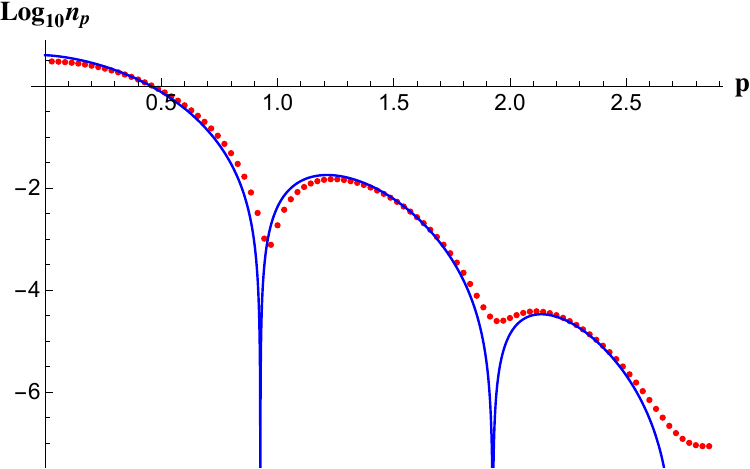}
  \end{minipage}
  \begin{minipage}[b]{0.55\linewidth}
    \centering
    \includegraphics[keepaspectratio, scale=0.5]{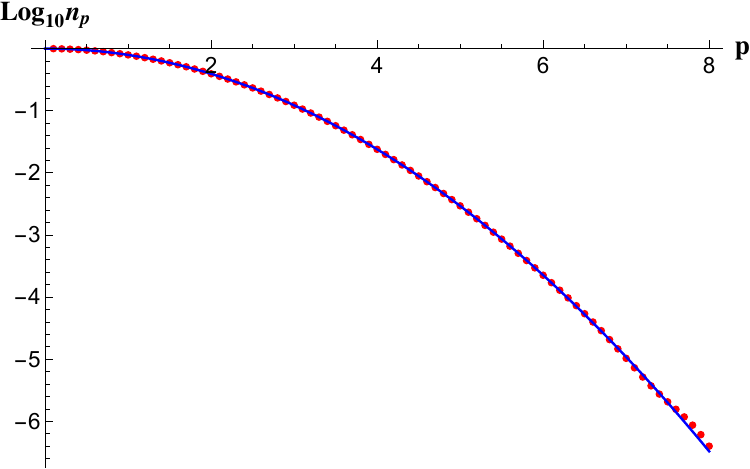}
  \end{minipage}
  \caption{Comparison between the analytic formula~\eqref{phin} (blue lines) and numerical results (red dots). The left panel shows the ${\bm r=\bm 1}$-case and the right the ${\bm r=\bm 1'}$-case. We see a good agreement for both cases.}
  \label{fig:phincomp}
\end{figure}

We consider the case that $\theta$ is fixed at $\theta=-\frac{M_{\rm pl}}{2\sqrt2}$ and parametrize $\phi$ as 
\begin{equation}
    \phi=\frac{M_{\rm pl}\log\left(\frac{\sqrt3}{2}\right)}{\sqrt 2}+v(t-t_0),
\end{equation}
with which the effective mass can be rewritten as
\begin{align}
    M_{\rm eff}^2=& \frac{81|C|^2}{64} m^2 \left(e^{\frac{\sqrt2 v(t-t_0)}{M_{\rm pl}}}-1\right)^4+\cdots \nonumber\\
    \approx&\frac{81|C|^2}{16} m^2\left(\frac{v(t-t_0)}{M_{\rm pl}}\right)^4\quad (\text{for ${\bm r = \bm 1}$}),\\
    M_{\rm eff}^2=&\frac{81|D|^2 m^2}{4} \left(e^{\frac{\sqrt2 v(t-t_0)}{M_{\rm pl}}}-1\right)^2+\cdots\nonumber\\
    \approx&\frac{81|D|^2 m^2}{2}\left(\frac{v(t-t_0)}{M_{\rm pl}}\right)^2\quad (\text{for ${\bm r = \bm 1'}$}),
\end{align}
where we have expanded the terms higher order in $v(t-t_0)/M_{\rm pl}$ since it is generally small. We notice that these expressions are the same as the $\phi$-fixed case~\eqref{1Dtheta1}, \eqref{1Dtheta1p} by replacing $|C|^2\to \frac{9}{16}|C|^2$ and $|D|^2\to \frac{3}{4}|D|^2$, respectively. Thus, the particle number density of the $k$-mode is given by
\begin{align}
    \Delta n_k=\Biggl\{\begin{array}{c}2e^{-\frac{2}{\sqrt3}\gamma_k}\left(1+\cos\frac{2}{\sqrt3}\gamma_k\right)\quad (\text{for ${\bm r = \bm 1}$})\\
    \exp \left[-\frac{2\pi k^2M_{\rm pl}}{9\sqrt2 |D|mv}\right]\quad (\text{for ${\bm r = \bm 1'}$}).
    \end{array}\label{phin}
\end{align}
Therefore, the effective force on $\phi$ due to particle production can be written as\footnote{Recall that $R\propto |C|^2$ and also that $F_\phi$ does not have an extra factor $({\rm Im}\tau)^2$ while $F_\theta$ does.}
\begin{align}
    F_\phi(t)\approx\frac{18\sqrt R a^3(t_0)(\partial_\phi M_{\rm eff}^2)(t)}{\pi \Gamma^4(\frac14)M_{\rm eff}(t)a^3(t)}\Theta(t-t_0) \quad (\text{for ${\bm r = \bm 1}$}),
\end{align}
or 
\begin{align}
    F_\phi(t)\approx \frac{a^3(t_0)(\partial_\phi M_{\rm eff}^2)(t)}{8\pi^3 a^3(t)M_{\rm eff}(t)}\left(\frac{9|D|mv}{\sqrt2 M_{\rm pl}}\right)^{\frac32}\Theta(t-t_0) \quad (\text{for ${\bm r = \bm 1'}$}).\label{Fphi1p}
\end{align}
We note that if there are $N$ fields that has the same Yukawa coupling, the effective force would be multiplied by $N$, which enhances the effect. We have checked the agreement between analytic formulas~\eqref{phin} and numerical results shown in Fig.~\ref{fig:phincomp}.

\begin{figure}[htbp]
  \begin{minipage}[b]{0.55\linewidth}
    \centering
    \includegraphics[keepaspectratio, scale=0.5]{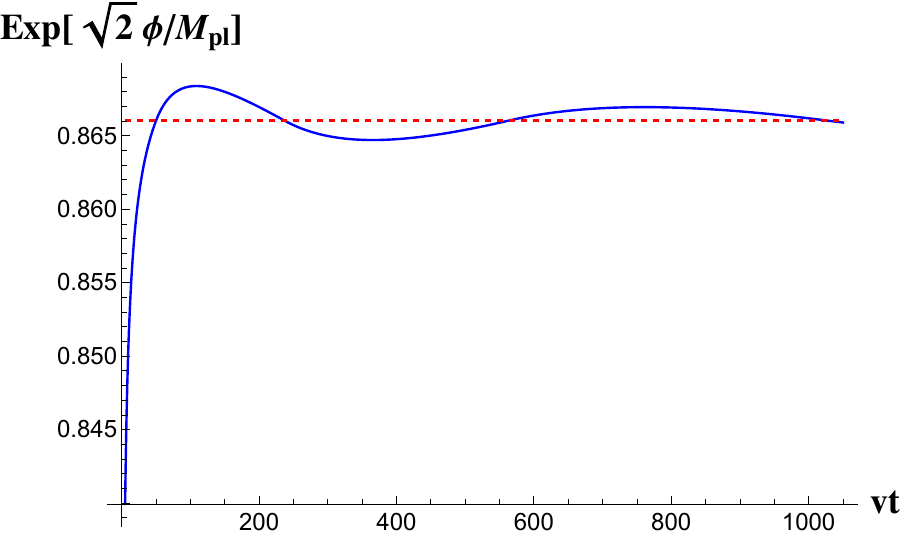}
  \end{minipage}
  \begin{minipage}[b]{0.55\linewidth}
    \centering
    \includegraphics[keepaspectratio, scale=0.5]{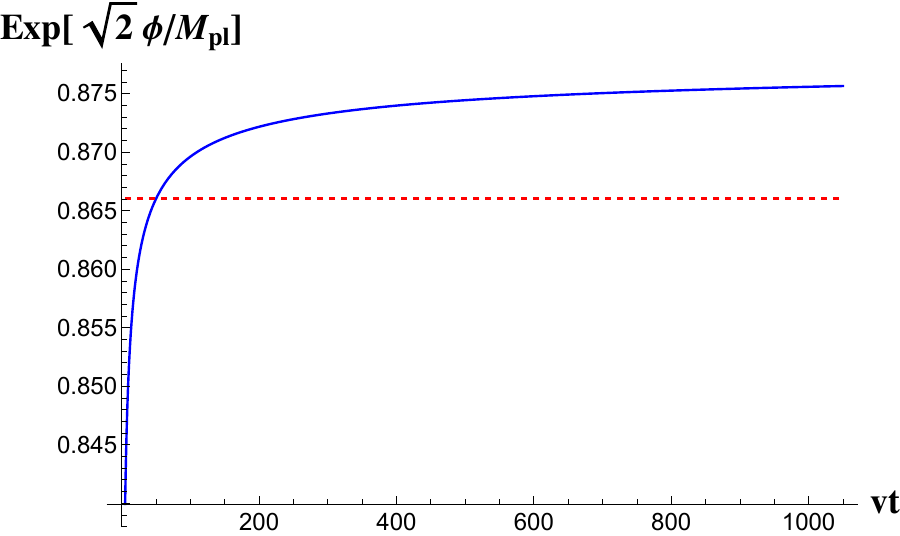}
  \end{minipage}
  \caption{A numerical solution of the E.O.M. of $\phi$ with (left) or without (right) particle production effect.  Here we have taken the following parameters: The expansion parameter $\beta=\frac12$, the initial time $\sqrt{v}t_{\rm ini}=5$, the time of particle production $\sqrt{v}t_0=50$, the Yukawa coupling parameter $m/\sqrt{v}=100$, and the species number to be $N=1$. The field velocity at the particle production is $\sqrt{v}=10^{-4}M_{\rm pl}$. The blue solid curve is the trajectory of $\phi$ and the red dashed line is the critical point ${\rm Im}\tau=\frac{\sqrt3}{2}$.}
  \label{fig:phi1p}
\end{figure}

\begin{figure}[htbp]
  \begin{minipage}[b]{0.55\linewidth}
    \centering
    \includegraphics[keepaspectratio, scale=0.5]{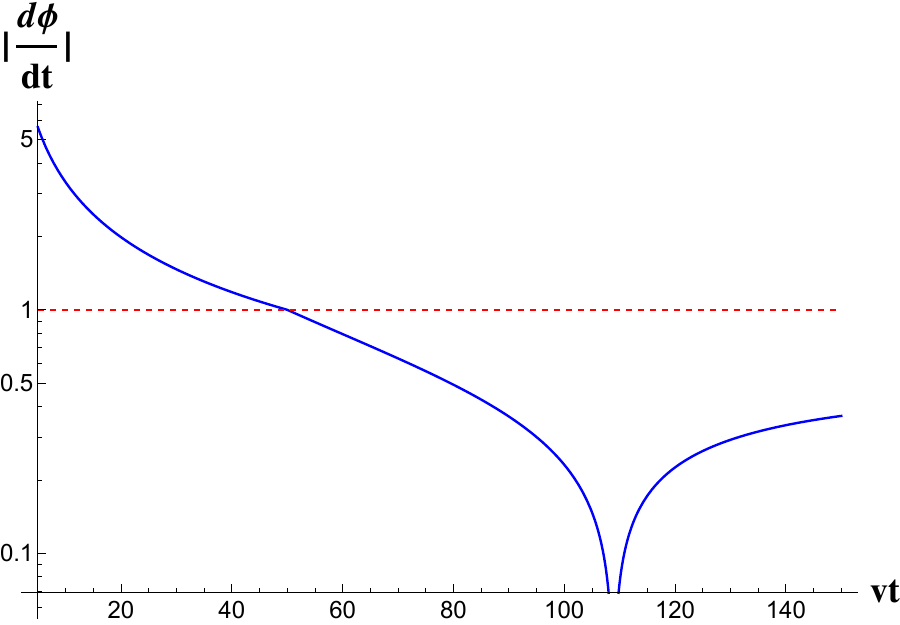}
  \end{minipage}
  \begin{minipage}[b]{0.55\linewidth}
    \centering
    \includegraphics[keepaspectratio, scale=0.5]{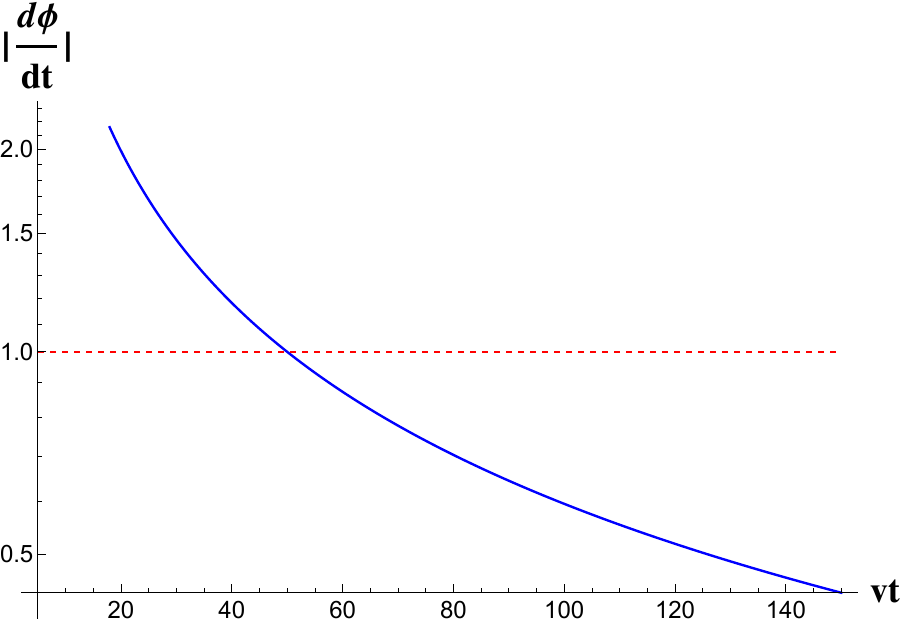}
  \end{minipage}
  \caption{The time dependence of the field velocity $\dot\phi$ with (left) or without (right) particle production effect. We have taken $v$ as the unit of the velocity. The parameters are the same as that in Fig.~\ref{fig:phi1p}. We see that in both cases the velocity is $1$ at $vt=vt_0=50$.}
  \label{fig:dphi1p}
\end{figure}

We show a numerical example of the moduli dynamics by using our analytic approximation in Fig~\ref{fig:phi1p}. In the example, we have used the ${\bm 1'}$-model. We have used the formulas~\eqref{freevelocity},\eqref{freeposition} to choose appropriate initial conditions such that $\phi$ crosses the ESP at $t_0$.\footnote{Another technical note: We have modified the denominator of \eqref{Fphi1p} as $M_{\rm eff}\to\sqrt{10^{-6}v+M_{\rm eff}^2}$ such that the singular behavior at the critical point is avoided. We have checked that the result is not affected by the IR cut-off. } As a cross-check of our numerical solution, we show the behavior of the field velocity $\dot\phi$ in Fig~\ref{fig:dphi1p}. As is clear from this simulation, the moduli trapping works despite Planck suppressed couplings between the matter field $\chi$ and the moduli. Thus, the moduli fields seem to prefer the ESP if they cross such a point along their time-evolution.

We emphasize that here only the first particle production event is taken into account, but as we see, there would be secondary and more particle production events when the modulus crosses the ESP, which further strengthen the trapping effect as quoted before.

\section{Summary and discussion}\label{summary}
In this work, we have studied the moduli trapping mechanism due to the particle production near ESPs within modular flavor symmetric models. We have reviewed and developed a general method for numerical simulations of models that the classical background fields and quantum fields interact with each other. Despite generality and simplicity, we have found that such an approach is not suitable for the application to modular flavor symmetric models because of the complexity of the couplings between background fields and quantum fields. Therefore, we have developed a semi-analytic approach where we analytically evaluate the particle production near the first crossing with the ESP, which yields the effective potential arising from produced particles. Although this approach looses the effect of subsequent particle production events, we have found the expected behavior, namely that the moduli in the modular flavor symmetric models can be trapped around the ESP at which there is a residual discrete symmetry. 

Although we have studied moduli trapping effects due to the scalar field $\chi$, 
we could discuss effects due to spinor and vector fields in a similar way.
As illustrating models, we have used the $A_4$ modular forms of weight 8, which have suppressed 
values around $\tau=\omega$.
Similar results would be obtained for modular forms of generic weights and other finite groups, 
if they have suppression behavior around $\tau=\omega$ and $\tau=\ri$.
Furthermore, concrete modular flavor symmetric models include many fields, whose masses are determined by similar modular forms.
Thus several modes would be produced around fixed points, which enlarges the effect of moduli trapping.

There are several issues that should be addressed in future work. One is to embed this mechanism into more realistic models of particle physics. In particular, it would be interesting to study the moduli dynamics within magnetized orbifold models where the standard model flavor structure including flavor mixings as well mass hierarchies has been realized~\cite{Abe:2012fj,Abe:2014vza,Fujimoto:2016zjs,Buchmuller:2017vho,Buchmuller:2017vut,Kikuchi:2021yog,Hoshiya:2022qvr}. It is also important to notice that the moduli trapping due to particle production never completes the moduli stabilization since the effective potential disappears as the particle numbers are diluted by expansion of the Universe, which therefore requires the introduction of additional potential terms. In particular, we have not taken into account the 1-loop effective potential such as the Coleman-Weinberg type potential that we have shown. Although it becomes small in supersymmetric models, it would be important to investigate the effect of such potential and the fate of moduli in the late Universe.

Another interesting issue would be the cosmological implications of moduli dynamics. It has been known that the particle production during inflation may give imprints on the curvature perturbation spectrum that can be seen from observations of cosmic microwave background~\cite{Green:2009ds,Barnaby:2009mc,Pearce:2016qtn,Flauger:2016idt,Pearce:2017bdc}. In particular, moduli in modular flavor symmetric models may become inflaton directions. (See e.g.~\cite{Kobayashi:2016mzg,Abe:2023ylh}.) Furthermore, the presence of the dynamically changing CP phase for the standard model matter fields may realize baryogenesis. In such a case, the method we have applied in this work would be useful to discuss the dynamics of matter and moduli simultaneously. We leave these interesting questions and model buildings for future work.

\section*{Acknowledgement}
This work was supported by JSPS KAKENHI Grant Numbers JP22KJ0047 (SK) and JP23K03375 (TK), 
JST SPRING Grant Number JPMJSP2119 (KN), and Waseda University Grant for Special Research Projects (Project number: 2023C-584) (YY).

\newpage
\appendix
\section{Review of particle production in time-dependent backgrounds}\label{PPreview}
We give a short review of particle production in time-dependent backgrounds. Let us consider a scalar field having a time-dependent mass,
\begin{align}
    \hat{\chi}(x)=\int \frac{d^3k}{(2\pi)^{\frac32}}\left[\hat{a}_{\bm k}e^{+\ri {\bm k}\cdot {\bm x}}f_{k}(t)+\hat{a}_{\bm k}^\dagger e^{-\ri {\bm k}\cdot {\bm x}}f_{k}^*(t)\right],
\end{align}
where $f(t)$ satisfies a mode equation~\eqref{modeeq} with $a(t)=1$ and the normalization condition~\eqref{KKnorm}. We have ``formally'' introduced creation and annihilation operators, which satisfies the canonical commutation relation~\eqref{CCR}. This is yet insufficient to give any meanings to the vacuum state that is annihilated by $\hat{a}_{\bm k}$ since we have not determined $f_k(t)$. More precisely speaking, unless the boundary condition of $f_k(t)$ is specified, the above expansion has no physical meaning. In Minkowski spacetime, the annihilation operators are introduced as coefficients of positive frequency modes $\sim \frac{1}{\sqrt{2\omega_k}}e^{-\ri \omega_k t}$. In time-dependent backgrounds, it is impossible to define a ``global'' positive frequency mode, but locally (in time) it is possible to find an approximate solution. For instance, the formal solution~\eqref{adiabatic} becomes asymptotically a positive frequency mode if $\alpha_k(t)\to 1$ and $\beta_k(t)\to0$ as $t\to -\infty$.\footnote{We assume that the background fields asymptote to be time-independent in the asymptotic past and future.} Indeed, under this condition, the scalar operator becomes
\begin{align}
    \hat{\chi}(x)\underset{t\to-\infty}{\to}\int \frac{d^3k}{(2\pi)^{\frac32}}\left[\hat{a}_{\bm k}e^{-\ri\omega_kt+\ri {\bm k}\cdot {\bm x}}+\hat{a}_{\bm k}^\dagger e^{+\ri\omega_kt-\ri {\bm k}\cdot {\bm x}}\right],
\end{align}
and the vacuum state that satisfies $\hat{a}_{\bm k}|0\rangle_{\rm in}=0$ is understood as a vacuum state in the asymptotic past. Thus we have defined a past (adiabatic) vacuum.

However, the ``past'' vacuum state is not a ``vacuum'' for a future observer. In general, $\beta_k(t)$ becomes non-zero due to Stokes phenomena\footnote{Review of Stokes phenomena can be found e.g. in \cite{Dumlu:2010ua,Dabrowski:2014ica,Dabrowski:2016tsx,Enomoto:2020xlf,Taya:2020dco,Hashiba:2021npn,Yamada:2021kqw}. For now, we just assume Stokes phenomena occur at some time and we are considering the time sufficiently far from the event.}, $t$ being asymptotic future time. In such a case, it is not appropriate to call $f_k(t)$ as a ``positive frequency mode''. Notice that the scalar operator can be expanded as
\begin{align}
    \hat{\chi}(x)&\underset{t\to+\infty}{\to}\int \frac{d^3k}{(2\pi)^{\frac32}}\left[(\alpha_k(\infty)\hat{a}_{\bm k}+\beta_k^*(\infty)\hat{a}^\dagger_{\bm k})e^{-\ri\omega_kt+\ri {\bm k}\cdot {\bm x}}+(\alpha_k^*(\infty)\hat{a}_{\bm k}^\dagger+\beta_k(\infty)\hat{a}_{\bm k}) e^{+\ri\omega_kt-\ri {\bm k}\cdot {\bm x}}\right]\nonumber\\
    &\equiv \int \frac{d^3k}{(2\pi)^{\frac32}}\left[\hat{b}_{\bm k}e^{-\ri\omega_kt+\ri {\bm k}\cdot {\bm x}}+\hat{b}^\dagger_{\bm k} e^{+\ri\omega_kt-\ri {\bm k}\cdot {\bm x}}\right].
\end{align}
The new set of creation and annihilation operators $\hat{b}_{\bm k},\hat{b}_{\bm k}^\dagger$ defines a future vacuum state $|0\rangle_{\rm out}$. Now, we find that the future creation and annihilation operators are given by linear combinations of the past creation annihilation operators. Note that there is no momentum exchange and the linear combination is diagonal with respect to momenta $\bm k$. 

How can we find particle production from vacuum? We define the future particle number density as
\begin{align}
    \hat{N}_k^{f}\equiv \hat{b}_{\bm k}^\dagger\hat{b}_{\bm k}=(\alpha_k^*(\infty)\hat{a}_{\bm k}^\dagger+\beta_k(\infty)\hat{a}_{\bm k})(\alpha_k(\infty)\hat{a}_{\bm k}+\beta_k^*(\infty)\hat{a}^\dagger_{\bm k}).
\end{align}
We would like to know how much ``future'' particles are contained in the past vacuum state~$|0\rangle_{\rm in}$, which can be explicitly evaluated as
\begin{align}
    {}_{\rm in}\langle 0|\hat{N}_k^f|0\rangle_{\rm in}=|\beta_k(\infty)|^2\langle\hat{a}_{\bm k}\hat{a}_{\bm k}^\dagger\rangle_{\rm in}=|\beta_k(\infty)|^2\frac{V}{(2\pi)^3}
\end{align}
where we have used $\lim_{\bm k\to \bm 0}\delta^3({\bm k})=\frac{V}{(2\pi)^3}$ and $V$ being spatial volume. Thus, we have found that the number density for (future) ${\bm k}$-mode is given by
\begin{align}
    n_{k}=|\beta_k(\infty)|^2.
\end{align}
It turns out that the past ``vacuum'' state is not a ``vacuum'' for a future observer viewpoint. The ``particle production from vacuum'' would be more appropriate to be understood as the ambiguity of energy, which does not allow us to globally define positive and negative modes.

We emphasize that the definition of ``particle'' is quite ambiguous except for constant backgrounds. In particular, when time-dependence is not turned off, we have to introduce adiabatic solutions such as WKB solutions, but there are infinitely many choices of adiabatic solutions. There is no clear answer to the question ``which solution should we take?'' but it is known that the optimal definition of the ``adiabatic particle number'' is related to the Stokes phenomena~\cite{Dabrowski:2014ica,Dabrowski:2016tsx}.\footnote{The optimal truncation of the adiabatic series was originally studied in \cite{Berry1,Berry2,Berry3}.}

\section{Modular forms}\label{modular-form}
The modular forms consist of the log derivatives of Dedekind eta function, which is given by
\begin{equation}
\eta(\tau)=q^{\frac{1}{24}}\prod_{n=1}^\infty (1-q^n),\quad q=e^{2\pi \ri \tau}.
\end{equation}
The $A_4$ modular forms of weight 2 are \cite{Feruglio:2017spp}
\begin{align}
    Y_1(\tau)=\frac{\ri}{2\pi}\left(\frac{\eta'(\tau/3)}{\eta(\tau/3)}+\frac{\eta'((\tau+1)/3)}{\eta((\tau+1)/3)}+\frac{\eta'((\tau+2)/3)}{\eta((\tau+2)/3)}-\frac{27\eta'(3\tau)}{\eta(3\tau)}\right),\\
    Y_2(\tau)=\frac{-\ri}{\pi}\left(\frac{\eta'(\tau/3)}{\eta(\tau/3)}+\omega^2\frac{\eta'((\tau+1)/3)}{\eta((\tau+1)/3)}+\omega\frac{\eta'((\tau+2)/3)}{\eta((\tau+2)/3)}\right),\\
    Y_3(\tau)=\frac{-\ri}{\pi}\left(\frac{\eta'(\tau/3)}{\eta(\tau/3)}+\omega\frac{\eta'((\tau+1)/3)}{\eta((\tau+1)/3)}+\omega^2\frac{\eta'((\tau+2)/3)}{\eta((\tau+2)/3)}\right),
\end{align}
where $\omega=e^{\ri \frac{2\pi}{3}}$. 
They correspond to the $A_4$ triplet.
By noting the fact that any positive integers can be written as $3m,3m-1,3m-2$, ($m\in\mathbb N$) and they consist of the logarithmic derivatives of the Dedekind eta function, one can easily rewrite these functions as
\begin{align}
    Y_1(\tau)=&1+\sum_{m=1}^\infty\left[\frac{9m q^m-18m^{2m}+9mq^{3m}}{(1-q^m)^3}+\frac{3(3m-2)q^{3m-2}}{1-q^{3m-2}}+\frac{3(3m-1)q^{3m-1}}{1-q^{3m-1}}-\frac{27q^{3m}}{1-q^{3m}}\right],\\
    Y_2(\tau)=&-6q^{-\frac23}\sum_{m=1}^\infty\left[\frac{(3m-2)q^m}{1-q^{3m-2}}+\frac{(3m-1)q^2m}{1-q^{3m-1}}\right],\\
    Y_3(\tau)=&-6q^{-\frac13}\sum_{m=1}^\infty\left[\frac{(3m-1)q^m}{(1-q^{3m-1}}+\frac{(3m-2)q^{2m-1}}{1-q^{3m-2}}\right].
\end{align}
We will approximate the infinite sum by some (sufficiently large) finite one. In our numerical simulations, we truncated the series up to some finite order.

By tensor products of $Y_1(\tau), Y_2(\tau)$, and $Y_3(\tau)$, 
we can write modular forms of higher weights.
The modular forms of weight 8 corresponding to 
three $A_4$ singlets, ${\bf 1}, {\bf 1}'$, and ${\bf 1}''$ can be written by \cite{Zhang:2019ngf}
\begin{align}
    Y^{(8)}_{\bf 1}(\tau) = (Y_1^2+2Y_2Y_3)^2, \quad
     Y^{(8)}_{{\bf 1}'}(\tau) = (Y_1^2+2Y_2Y_3)(Y_3^2+2Y_1Y_2),  \quad
       Y^{(8)}_{{\bf 1}''}(\tau) = (Y_3^2+2Y_1Y_2)^2. 
\end{align}

\section{Field dynamics without potential in expanding Universe}\label{freedynamics}
Here we discuss the free field dynamics in the expanding Universe. The E.O.M. of a massless free scalar field in the FRW background is given by
\begin{equation}
    \ddot{\phi}+3H\dot\phi=0,
\end{equation}
which can also be written as
\begin{equation}
    \frac{1}{a^3}\frac{d}{dt}\left(a^3\dot\phi\right)=0.
\end{equation}
Therefore, the first integration yields
\begin{align}
    \dot\phi(t)=\frac{a^3(t_r)}{a^3(t)}\dot\phi(t_r)=\left(\frac{t_r}{t}\right)^{3\beta}\dot\phi(t_r),\label{freevelocity}
\end{align}
where $t_r$ is a reference time. The secondary time integration leads to
\begin{align}
    \phi(t)=\phi(t_r)+\frac{1}{1-3\beta}\left[\left(\frac{t_r}{t}\right)^{3\beta-1}-1\right]t_r\dot\phi(t_r),\label{freeposition}
\end{align}
assuming $\beta\neq\frac13$. Near $t=t_r$, we may expand the above expression as
\begin{align}
    \phi(t)=\phi(t_r)+\dot\phi(t_r)(t-t_r)-\frac{3\beta \dot\phi(t_r)(t-t_r)^2}{2t_r}+\mathcal{O}((t-t_r)^3).
\end{align}
Now since $H(t_r)\sim 1/t_r$, we can neglect the third and higher order terms as long as Hubble parameter at $t=t_r$ is sufficiently smaller than $\sqrt{|\dot\phi(t_r)|}$. 

The relation between the initial velocity $v_{\rm ini}=\dot\phi(t_{\rm ini})$ and $v=\dot\phi(t_0)$ at some reference time $t_0>t_{\rm ini}$ is 
\begin{equation}
    v=\left(\frac{t_{\rm ini}}{t_0}\right)^{3\beta}v_{\rm ini}.
\end{equation}
This would be useful to relate the initial velocity with the field velocity at the crossing with enhanced symmetry points particularly in estimating the particle production rate.

\bibliographystyle{JHEP}
\bibliography{main-2.bib}
\end{document}